\documentclass{LMCS}
\newtheorem{theorem}{Theorem}
\newtheorem{lemma}{Lemma}

\newtheorem{claim}{Claim}

\newtheorem{proposition}{Proposition}

\theoremstyle{definition}
\newtheorem{definition}{Definition}
\newtheorem{example}{Example}

\def\endproof{\qed}

\long\def\BEGINCOMMENT #1\ENDCOMMENT{\relax}

\newcommand{\maps}\longrightarrow
\newcommand{\cmaps}\Longrightarrow
\newcommand{\vocab}{\tau}

\newcommand{\pest}{\operatorname{{\bf P}}}

\newcommand{\dest}{\operatorname{{\bf D}}}
\newcommand{\cest}{\operatorname{{\bf C}}}
\newcommand{\best}{\operatorname{{\bf B}}}
\newcommand{\aest}{\operatorname{{\bf A}}}
\newcommand{\kest}{\operatorname{{\bf K}}}
\newcommand{\hest}{\operatorname{{\bf H}}}
\newcommand{\gest}{\operatorname{{\bf G}}}
\newcommand{\csp}{\operatorname{CSP}}

\newcommand{\arity}{\rho}

\newcommand{\imply}[2]{{\bf (#1)$\rightarrow$(#2)}}

\newcommand{\comp}{\circ}

\newcommand{\canonic}{\Theta}

\newcommand{\dom}{\operatorname{dom}}
\newcommand{\str}{\operatorname{STR}}

\newcommand{\infinitary}{L_{\infty,\omega}}
\newcommand{\infinitarybounded}{L_{\infty,\omega}^\omega}
\newcommand{\pathlogic}{M}
\newcommand{\restpathlogic}{N}
\newcommand{\conjrestpathlogic}{O}

\def\doi{1 (1:5) 2005}
\lmcsheading%
{\doi}
{32}  %% no. of pages
{}
{}
{Sep.~22, 2004} %% only use ``\phantom{0}'' for single digit dates!
{Apr.~29, 2005} %% only use ``\phantom{0}'' for single digit dates!
{}    %% date of revision, if applicable

\begin{document}

\title{Linear Datalog and Bounded Path Duality of Relational Structures}

\author[V.~Dalmau]{V\'{\i}ctor Dalmau}
\thanks{Research conducted whilst the author was visiting the University of California, Santa Cruz, supported
by NSF grant CCR--9610257}
\address{Departament de Tecnologia, Universitat Pompeu Fabra, \\
Estaci\'o de Fran\c{c}a, Passeig de la Circumval.lacio 8. Barcelona 08003, Spain \\
Fax: +34 93 542 24 49}
\email{victor.dalmau@tecn.upf.es}

 
\begin{abstract}

In this paper we systematically investigate the connections between
logics with a finite number of variables, structures of bounded
pathwidth, and linear Datalog Programs. We prove that, in the context
of Constraint Satisfaction Problems, all these concepts correspond to
different mathematical embodiments of a unique robust notion that we
call bounded path duality.  We also study the computational complexity
implications of the notion of bounded path duality.  We show that
every constraint satisfaction problem $\csp(\best)$ with bounded path
duality is solvable in NL and that this notion explains in a uniform
way all families of CSPs known to be in NL.  Finally, we use the
results developed in the paper to identify new problems in NL.

\end{abstract}

\maketitle

\section{Introduction}

The constraint satisfaction problem provides a framework in which it
is possible to express, in a natural way, many combinatorial problems
encountered in artificial intelligence and elsewhere. A constraint
satisfaction problem is represented by a set of variables, a domain of
values for each variable, and a set of constraints between variables.
The aim of a constraint satisfaction problem is then to find an
assignment of values to the variables that satisfies the constraints.

Solving a general constraint satisfaction problem is known to be
NP-complete~\cite{Cook:1971,Mackworth:1977}.  One of the main
approaches pursued by researchers in artificial intelligence and
computational complexity to tackle this problem has been the
identification of tractable cases obtained by imposing restrictions in
the constraints
(see~\cite{Cooper/Cohen/Jeavons:1994,Dalmau:2000,Dalmau/Pearson:1999,Dechter/Pearl:1988,Feder/Vardi:1998,Freuder:1985,Jeavons/Cohen/Cooper:1998,Jeavons/Cohen/Gyssens:1997,Kirousis:1993,Montanari:1974,Montanari/Rossi:1991,Beek/Dechter:1995,Hentenryck/Deville/Teng:1992}).

Recently~\cite{Feder/Vardi:1998} (see also~\cite{Jeavons:1998}) it has
been observed that the constraint satisfaction problem can be recast
as the following fundamental algebraic problem: given two finite
relational structures $\aest$ and $\best$, is there a homomorphism
from $\aest$ to $\best$? In this framework, the problem of identifying
which restrictions in the constraints guarantee tractability is
equivalent to deciding for which structures $\best$, the homomorphism
problem when only $\aest$ is part of the input, denoted by
$\csp(\best)$, is solvable in polynomial time. In this paper we study
this framework.

One of the simplest examples of constraint satisfaction problems of
this form is obtained when $\best$ is fixed to be a $k$-clique
$\kest_k$ for some $k\geq 2$. In this case, an instance
$\csp(\kest_k)$ is again a graph $\aest$ and the question of existence
of an homomorphism from $\aest$ to $\best$ is equivalent to deciding
whether $\aest$ is $k$-colorable.

Feder and Vardi~\cite{Feder/Vardi:1998} introduced one general
condition for tractability of $\csp(\best)$ that accounts for many of
the tractable cases of $\csp(\best)$. More precisely, Feder and Vardi
observed that for many polynomial-time solvable Constraint
Satisfaction Problems of the form $\csp(\best)$ there is a {\em
Datalog Program} that defines the complement of $\csp(\best)$.

In order to illustrate this consider now the particular case when
$\best$ is $\kest_2$, that is, $\best$ is a graph with two nodes and
an edge between them. As it has been observed before, $\csp(\kest_2)$
is the set of all $2$-colorable graphs. It is well-known that a graph
is $2$-colorable iff it does not have odd cycles. The following
Datalog Program asserts that the graph $\aest=(V,E)$ contains an odd
cycle:

$$\begin{array}{rl}
P(x,y) \; \text{:--} & E(x,y) \\
P(x,y) \; \text{:--} & P(x,z), E(z,u), E(u,y) \\
Q \; \text{:--} & P(x,x) 
\end{array}$$

This Datalog Programs has three rules. These rules are a recursive specification of two predicates $P$, and $Q$, called IDBs. 
Predicate $P(x,y)$ holds whether there exists a path in $\aest$ of odd length from $x$ to $y$ and predicate $Q$, which acts
as goal predicate, holds if there exists a cycle of odd length . 

Since the seminal results of Feder and Vardi, the language Datalog has played a prominent 
role in the study of the complexity of the CSP. In particular, several connections between Datalog Programs and
well-established notions developed in the area of constraint satisfaction problems and graph homomorphism
have been explored.

One of this notions is that of bounded treewidth duality initially
introduced in the area of ${\bf
H}$-coloring~\cite{Hell/Nesetril/Zhu:1996}, which can be reformulated
as the constraint satisfaction problem $\csp(\hest)$ where $\hest$ is
a graph.  In a $\hest$-coloring problem we are given a graph $\gest$
as an input and we are asked where there exists an homomorphism from
$\gest$ to $\hest$. It has been observed~\cite{Hell/Nesetril/Zhu:1996}
that the vast majority of the tractable cases of $\csp(\hest)$ have an
obstruction set of bounded treewidth. An obstruction set of
$\csp(\hest)$ is any set $S$ of graphs not homomorphic to $\hest$ such
that for every graph $\gest$ not homomorphic to $\hest$ there exist a
graph in $S$ that is homomorphic to $\gest$. Every graph $\hest$
having an obstruction set of bounded treewidth is said to have bounded
treewidth duality. Bounded treewidth duality turns out to be
equivalent to definability in Datalog, as shown
in~\cite{Feder/Vardi:1998}. More precisely, a graph $\hest$, has
bounded treewidth duality if and only if the complement of
$\csp(\hest)$ is definable in Datalog.  The relationship between
Constraint Satisfaction Problems, Datalog and structures of bounded
treewidth has been further investigated
in~\cite{Kolaitis/Vardi:2000a},~\cite{Kolaitis/Vardi:2000b}, and
~\cite{Dalmau/Kolaitis/Vardi:2002}.

The goal of the present paper is to take a closer look inside the
structure of Datalog Programs. In particular, we are interested in
{\em linear} Datalog Programs~\cite{Abiteboul/Hull/Vianu:1995}, which
are Datalog Programs in which every rule has at most one IDB in the
body. Many constraint satisfaction problems solvable with a Datalog
Program are indeed solvable by a linear Datalog Program. In fact, the
Datalog Program defining NON-2-COLORABILITY presented above is a
linear Datalog Program.  It is well known that problems solvable by a
linear Datalog Program are solvable in non-deterministic logarithmic
space.  This class has received a lot of interest in complexity
theory. In particular, it is known that $\text{NL}\subseteq\text{NC}$
and, therefore, problems in NL are highly parallelizable.

In this paper we embark on a systematic study of the relationship
between linear Datalog Programs, finite variable logics and structures
of bounded pathwidth. We prove several different but equivalent
characterizations of definability in linear Datalog. Most of them are
in the realm of logic. In particular, we show that for every
constraint satisfaction problem of the form $\csp(\best)$,
$\neg\csp(\best)$ is definable in linear Datalog if and only if
$\csp(\best)$ is definable in the logic $M^{\omega}$, which is defined
to be the subset of the infinitary logic $\infinitary^{\omega}$ in
which only existential quantification, infinitary disjunction and a
certain restricted infinitary conjunction is allowed. Definability in
linear Datalog is also shown to be equivalent to definability in
restricted Krom SNP, which is defined to be the set of all existential
second order sentences with a universal first-order part in prenex CNF
form, in which we require every clause of the first-order part to
contain only negated relation symbols and at most one positive and one
negative second order-variable.  We also obtain a combinatorial
reformulation of definability in linear Datalog. More precisely, we
show that for every structure $\best$, the complement of $\csp(\best)$
is definable in linear Datalog if and only if $\csp(\best)$ has an
obstruction set of bounded pathwidth. All of these different but
equivalent reformulations of the same notion seem to provide some
evidence that the class of problems definable is linear Datalog is an
interesting and robust class.

Finally we use our results to investigate which constraints satisfaction problems are in NL.
Despite the large amount of tractable cases of constraint satisfaction problems identified so far, very few subclasses of constraint satisfaction problems
are known to be in NL. To our knowledge, the only families of CSP problems known to be in NL are the class of bijunctive satisfiability 
problems~\cite{Schaefer:1978}, including 2-SAT, which later on was generalized to the class of implicational constraints~\cite{Kirousis:1993} (see 
also~\cite{Jeavons:1998}), the class of implicative Hitting-Set Bounded (see~\cite{Creignou/Khanna/Sudan:2001}) originally defined (although with
a different name) in~\cite{Schaefer:1978}, and the class of posets with constants invariant under a near-unanimity operation~\cite{Krokhin/Larose:stacs03}. First we observe that all this families of problems are particular cases of bounded path duality problems. Consequently, the results in our paper provide an uniform explanation of all known constraint satisfaction problems known to be in NL. Finally,
we identify some new families of constraint satisfaction problems solvable in NL.

\section{Basic Definitions}

\BEGINCOMMENT
{\tt Result about the core and expressibility}
\ENDCOMMENT

Most of the terminology introduced in this section is fairly standard.
We basically follow~\cite{Grohe:03}. 
A {\em vocabulary} is a finite set of relation symbols or predicates.
 In the following $\vocab$ always denotes a vocabulary. Every
relation symbol $R$ in $\vocab$ has an {\em arity} $r=\arity(R)\geq 0$ 
associated to it. We also say that $R$ is an $r$-ary relation
symbol. 

A $\vocab$-structure $\aest$ consists of a set $A$, called 
the {\em universe} of $\aest$, and a relation $R^{\aest}\subseteq A^r$ 
for every relation symbol $R\in\vocab$ where $r$ is the arity of $R$.
Unless otherwise stated we will assume that we are dealing with
{\em finite} structures, i.e., structures with a finite universe. 
Throughout the paper we use the 
same boldface and slanted capital letters to denote a structure
and its universe, respectively.

Let $\aest$ and $\best$ be $\vocab$-structures. We say that $\best$ is
a substructure of $\aest$, denoted by $\best\subseteq\aest$, if
$B\subseteq A$ and for every $R\in\vocab$, $R^{\best}\subseteq
R^{\aest}$.  If $\aest$ is a $\vocab$-structure and $B\subseteq A$,
then $\aest_{|B}$ denotes the substructure induced by $\aest$ on $B$,
i.e., the $\vocab$-structure $\best$ with universe $B$ and
$R^{\best}=R^{\aest}\cap B^r$ for every $r$-ary $R\in\vocab$.

Let $\aest$ and $\best$ be $\vocab$-structures. We denote by $\aest\cup\best$
the $\vocab$-structure with universe $A\cup B$ and such that for
all $R\in\vocab$, $R^{\aest\cup\best}=R^{\aest}\cup R^{\best}$.

A {\em homomorphism} from a $\vocab$-structure $\aest$ to a
$\vocab$-structure $\best$ is a mapping $h:A\rightarrow B$ such that
for every $r$-ary $R\in\vocab$ and every $\langle
a_1,\dots,a_r\rangle\in R^{\aest}$, we have $\langle
h(a_1),\dots,h(a_r)\rangle\in R^{\best}$. We denote this by
$\aest\stackrel{h}\maps\best$.  We say that $\aest$ homomorphically
maps to $\best$, and denote this by $\aest\maps\best$ iff there exists
some homomorphism from $\aest$ to $\best$. We denote by
$\hom(\aest,\best)$ the set of all homomorphisms from $\aest$ to
$\best$.

We will assume by convention that for every set $B$ there
exists one mapping $\lambda:\emptyset\rightarrow B$. Consequently, if
$\aest$ is a structure with an empty universe then
$\{\lambda\}=\hom(\aest,\best)$.

Let $a_1,\dots,a_m$ be elements in $A$ and let $b_1,\dots,b_m$ be
elements in $B$. 
We shall write $\aest,a_1,\dots,a_m\maps
\best,b_1,\dots,b_m$ to denote that there exists some homomorphism $h$ from
$\aest$ to $\best$ such that $h(a_i)=b_i$, $1\leq i\leq m$.

Let $\aest$ be a $\vocab$ structure and let
$\vocab'\subseteq\vocab$. We denote by $\aest[\vocab']$ the
$\vocab'$-structure such that for every $R\in\vocab'$,
$R^{\aest[\vocab']}=R^{\aest}$. Similarly, if $\mathcal C$ is a
collection of $\vocab$-structures we denote by $\mathcal C[\vocab']$
the set $\{\aest[\vocab']:\aest\in\mathcal C\}$.  $\str$ denotes the
class of all structures and consequently, $\str[\vocab]$ denotes the
class of all $\vocab$-structures.

\BEGINCOMMENT {\tt Here I might want a add the Cartesian product
operation and later on introduce the lattice given by the equivalence
classes under homomorphism and say that over infinite lattices is
indeed an algebraic lattice

study whether to decide whether the lattice of finite structures is
algebraic is equivalent to THE problem } \ENDCOMMENT

Finally, $\csp(\best)$ is defined to be the set of all 
structures $\aest$ such that $\aest\maps\best$.

\BEGINCOMMENT
\section{Datalog Programs}

{\tt Maybe later}
\ENDCOMMENT

\section{Infinitary Logic}

The following definition is borrowed from~\cite{Kolaitis/Vardi:1995}.
Let $\vocab$ be a vocabulary containing only relational symbols and let $\{v_1,\dots,v_n,\dots\}$ be a
countable set of variables. The class $\infinitary$ of {\em infinitary
formulas} over $\vocab$ is the smallest collection of formulas such
that
\begin{itemize}
\item it contains all first-order formulas over $\vocab$.
\item if $\varphi$ is a formula of $\infinitary$ then so is $\neg\varphi$.
\item if $\varphi$ is a formula of $\infinitary$ and $v_i$ is a variable, then
$(\forall v_i)\varphi$ and $(\exists v_i)\varphi$ are also formulas of $\infinitary$.
\item if $\Psi$ is a set (possibly infinite) of $\infinitary$ formulas, then $\bigvee\Psi$ and $\bigwedge\Psi$
are also formulas of $\infinitary$.
\end{itemize}

The first subscript of $\infinitary$ indicates that conjunctions and disjunctions
can be taken over arbitrary infinite sets and the second that only finite quantifier
blocks are allowed. 

\BEGINCOMMENT
Remark: Unless otherwise explicitly stated, we do not allow the use of equalities (=) in
the formulas. This restriction is not essential and simplifies matters slightly. 
The main reason for this is that the notion of homomorphism, when applied over structures that
contain the relation equality, or some other built-in predicate, must be modified in order
to apply only to non-built-in predicates. Furthermore, we will be mainly interested in
classes of structures closed under homomorphism, for which equality does not add, in
general, expressiveness. 
\ENDCOMMENT

The concept of a {\em free variable} in a formula of $\infinitary$ is defined in the same way
as for first-order logic. We use the notation $\varphi(u_1,\dots,u_m,\dots)$ to denote that
$u_1,\dots,u_m,\dots$ are different and that
$\varphi$ is a formula of $\infinitary$ whose free variables are among the variables 
$u_1,\dots,u_m,\dots$. A {\em sentence} of $\infinitary$ is a formula $\varphi$ of $\infinitary$
with no free variables. The semantics of $\infinitary$ is a direct extension of the semantics
of first-order logic, with $\bigvee\Psi$ interpreted as a disjunction over all formulas in $\Psi$
and $\bigwedge\Psi$ interpreted as a conjunction. If $\aest$ is a structure over $\vocab$ and 
$a_1,\dots,a_m,\dots$ is a sequence of (not necessarily different) elements from the universe
of $\aest$, then we write
$$\aest,a_1,\dots,a_m,\dots,\models\varphi(u_1,\dots,u_m,\dots)$$
to denote that the structure $\aest$ {\em satisfies} the formula $\varphi$ of $\infinitary$ when
each variable $u_i$ is interpreted by the element $a_i$.

Let $k$ be a non-negative integer. The {\em infinitary logic with $k$ variables}, denoted by $\infinitary^k$ consists of all
formulas of $\infinitary$ with at most $k$ distinct variables.
\BEGINCOMMENT
The infinitary logic $\infinitarybounded$ consists of all formulas of $\infinitary$ with
a finite number of distinct variables. Thus,
$$\infinitarybounded=\bigcup_{k=1}^\infty\infinitary^k$$
\ENDCOMMENT
Let $L^k$ be the collection of all formulas of $\infinitary^k$, that are obtained from
atomic formulas using infinitary disjunctions, infinitary conjunctions, and existential quantification
only. 
\BEGINCOMMENT
We also put
$$L^\omega=\bigcup_{k=1}^\infty L^k$$
and call this logic the {\em existential negation-free fragment of $\infinitarybounded$}
\ENDCOMMENT

Let $j$ be a non-negative integer. We will denote by $j$-{\em restricted infinitary conjunction} the infinitary 
conjunction $\bigwedge\Psi$ when $\Psi$ is a collection of $\infinitary$ formulas
such that (a) every formula with more than $j$ free variables is quantifier-free and
(b) at most one formula in $\Psi$ having quantifiers is not a sentence. 
If furthermore the set $\Psi$ is finite then we will call it $j$-{\em 
restricted conjunction}.

Let $0\leq j\leq k$ be non-negative integers.
Let $\pathlogic^{j,k}$ ($\restpathlogic^{j,k}$) be the collection of all formulas of $\infinitary^k$,
that are obtained from atomic formulas using infinitary disjunction, 
$j$-restricted infinitary conjunction ($j$-restricted conjunction), and existential quantification only. 
Finally, let $\conjrestpathlogic^{j,k}$ be the collection of 
all formulas of $\infinitary^k$, 
that are obtained from atomic formulas using $j$-restricted conjunction, 
and existential quantification only. We also put 
$$\pathlogic^{\omega}=\bigcup_{0\leq j\leq k} \pathlogic^{j,k}, \; \;
\restpathlogic^{\omega}=\bigcup_{0\leq j\leq k} \restpathlogic^{j,k}, \; \;
\text{ and } \conjrestpathlogic^{\omega}=\bigcup_{0\leq j\leq k} \conjrestpathlogic^{j,k}$$

The following example illustrates the expressive power of these logics.

\begin{example}[Paths, Bipartiteness]

Assume that the vocabulary consists of a unique binary relation $E$, and let $\varphi_{n}(x,y)$, $n\geq 1$ be the first order
formula asserting that there exists a path or length $n$ from $x$ to $y$. The naive way to write $\varphi_{n}(x,y)$
requires $n+1$ variables, namely
$$\exists x_1\exists x_2\dots\exists x_{n-1}  E(x,x_1)\wedge E(x_1,x_2)\wedge\cdots\wedge E(x_{n-1},y)$$

It is well-known that $\varphi_{n}$ is equivalent to a formula in $\infinitary^3$ (in fact in $L^3$). To see this, put
$\varphi_1(x,y)\equiv E(x,y)$
and assume, by induction on $n$, that $\varphi_{n-1}(x,y)$ is equivalent to a formula in $\infinitary^3$. Then
$$\varphi_{n}(x,y)=\exists z [E(x,z)\wedge \exists x ((z=x)\wedge\varphi_{n-1}(x,y))]$$

A closer look at $\varphi_n(x,y)$ reveals that
every conjunction used in the definition of
$\varphi_n(x,y)$ is $j$-restricted and hence we
can conclude that $\varphi_{n}(x,y)$ is 
in $\conjrestpathlogic^{2,3}$ (and hence in $\restpathlogic^{2,3}$ and $\pathlogic^{2,3}$).

Finally, let ${\mathcal C}$ be the set of all finite $\vocab$-structures that, interpreted as graphs, are bipartite. 
It is well known that a graph is bipartite if and only if does not contain odd cycles. Therefore, the set 
${\mathcal C}$ is defined by the the formula 
$$\exists x \;\; \bigwedge_{n\geq 0} \varphi_{2n+1} (x,x)$$

\end{example}

As we shall see later (Theorem~\ref{the:gamesformulas}) the expressive power of this
logics is the same regardless on whether or not we allow the use of equalities (=) in the formulas. 
By convention we shall assume that, unless otherwise explicitly stated, formulas
do not contain equalities.

We finish this section by stating without proof a very simple fact about these logics that will be used
intensively in our proofs.

\begin{proposition}
\label{pro:hompressat}
Let $\aest$, $\best$ be $\vocab$-structures such that $\aest\maps\best$. 
For every sentence $\varphi$ in $\infinitary^{\omega}$ that
does not contain universal quantification nor negation such that $\aest\models\varphi$ we
have that $\best\models\varphi$.
\end{proposition}

\BEGINCOMMENT
We also put
$$\pathlogic^k=\bigcup_{0\leq j\leq k} \pathlogic^{j,k},$$
and
$$\pathlogic^\omega=\bigcup_{k=1}^\infty \pathlogic^k$$
\ENDCOMMENT

\BEGINCOMMENT
Let $0\leq j\leq k$ be non-negative integers.
Let $\restpathlogic^{j,k}$ be the collection of all formulas of $\infinitary^k$, 
that are obtained from atomic formulas using infinitary disjunction, 
$j$-restricted conjunction, and existential quantification only. 
\ENDCOMMENT
\BEGINCOMMENT
We also put
$$\restpathlogic^k=\bigcup_{0\leq j\leq k} \restpathlogic^{j,k},$$
and
$$\restpathlogic^\omega=\bigcup_{k=1}^\infty \restpathlogic^k$$
\ENDCOMMENT
\BEGINCOMMENT
Let $0\leq j\leq k$ be non-negative integers.
Let $\conjrestpathlogic^{j,k}$ be the collection of all formulas of $\infinitary^k$, 
that are obtained from atomic formulas using $j$-restricted conjunction, 
and existential quantification only. 
\ENDCOMMENT
\BEGINCOMMENT
We also put
$$\conjrestpathlogic^k=\bigcup_{0\leq j\leq k} \restpathlogic^{j,k},$$
and
$$\conjrestpathlogic^\omega=\bigcup_{k=1}^\infty \restpathlogic^k$$
\ENDCOMMENT

\section{Quasi-Orderings and Pathwidth of Relational Structures}

A {\em quasi-ordering} on a set $S$ is a reflexive and transitive relation $\leq$ on $S$.
Let $\langle S,\leq\rangle$ be a quasi-ordered set. Let $S',S''$ be subsets of $S$.
We say that $S'$ is a {\em filter} if it is closed under $\leq$ upward; that is, if $x\in S'$
and $x\leq y$, then $y\in S'$. 
The {\em filter generated} by $S''$ is the
set $F(S'')=\{ y\in S: \exists x\in S''\;  x\leq y\}$.
We say
that $S'$ is an {\em ideal} if it is closed under $\leq$ downward; that is, if $x\in S'$
and $y\leq x$, then $y\in S'$. The {\em ideal generated} by $S''$ is the
set $I(S'')=\{ y\in S: \exists x\in S''\; y\leq x\}$. Observe that every subset $S'$ of
$S$ and its complement $S\backslash S'$ satisfy the following relation: $S'$ is an 
ideal iff  $S\backslash S'$ is a filter.

Let $I$ be an ideal of $\langle S,\leq\rangle$.
We say that a set $O\subseteq S$ forms an {\em
obstruction set} for $I$ if $$x\in I \text{ iff } \forall y\in O (y\not\leq x)$$
That is, $O$ is an obstruction set for $I$ if $I$ is the complement of $F(O)$

Let $\vocab$ be a vocabulary. The set of $\vocab$-structures, $\str[\vocab]$, is quasi-ordered by
the homomorphism relation. In consequence, a set $C$ of $\vocab$-structures 
is an ideal of $\langle\str[\vocab],\maps\rangle$ if
$$\best\in C, \aest\maps\best \implies \aest\in C$$
Observe that for any relational structure $\best$, $\csp(\best)=I(\best)$.

Let us define a notion of pathwidth relative to relational structures, which is
the natural generalization of the notion of pathwidth over graphs, introduced
by Robertson and Seymour~\cite{Robertson/Seymour:MinorI}. We follow
the lines of previous generalizations of similar notions as treewidth.
For reasons that will be made clear later it is desirable to parameterize the
ordinary notion of pathwidth to capture a finer structure. For this
purpose we will consider not only the maximum size of any set of the path-decomposition
but also the maximum size of its pairwise intersection.

\begin{definition}
Let $\aest$ be a $\vocab$-structure. A {\em path-decomposition} of $\aest$ 
is a collection $S_1,\dots,S_n$, $S_i\subseteq A$ such that:
\begin{enumerate}
\item for every $r$-ary relation symbol $R$ in $\vocab$ and every $\langle a_1,\dots,a_r\rangle\in R^{\aest}$,
there exists $1\leq i\leq n$ such that $\{a_1,\dots,a_r\}\subseteq S_i$.
\item if $a\in S_i\cap S_j$, then $a\in S_l$ for all $i\leq l\leq j$.
\end{enumerate} 
The {\em width} of the path-decomposition is defined to be the pair 
$\langle\max\{|S_i\cap S_{i+1}|: 1\leq i\leq n-1\},\max\{|S_i|:1\leq i\leq n\}\rangle$.
We say that a structure $\aest$ has {\em pathwidth} at most $(j,k)$ if it has a path
decomposition of width $(j,k)$. 

The concept of pathwidth relative to
relational structures introduced here is intended to be a natural
generalization of the notion of pathwidth defined over graphs~\cite{Robertson/Seymour:MinorI}. However
there are some points in which our definition is not standard. First, in the ordinary
notion of pathwidth over graphs, as defined in~\cite{Robertson/Seymour:MinorI}, the 
width of a path-decomposition is defined as $\max\{|S_i|:1\leq i\leq n\}-1$. We are interested
in a more fine-grained analysis that motivates the consideration, not only of the cardinality
of the sets but also the cardinality of the intersection of two consecutive sets in the
intersection. Furthermore, it is convenient for us not to subtract $1$, as it
is customary, as then this subtraction would had to be carried over all the paper.
%The second aspect in which our notion of pathwidth differs from
%the usual notion of pathwidth is that we do not consider the minimum among all
%path-decompostions of a graph. That is, from the customary view point it is only
%valid to say that a graph $G$ has pathwidth $k$ if $k$ is the minimum width among all the
%path-decompositions of $G$, but according to our definition a if a graph has path width $(j,k)$
%then it also has pathwidth $(j',k')$ for any $j',k'$ with $j\leq j'$ and $k'\leq k$. 

In order to delineate even more the precise relationship between
the notion of pathwidth as it is usually defined over graphs and
the notion of pathwidth of relational structures introduced in this
paper we remark the following equivalence.
 
Let $\aest$ be a relational structure. The the following numbers are equal:
\begin{itemize}
\item the pathwidth of the Gaifman graph of $\aest$ plus one.
\item the minimum $k$ such that $\aest$ has pathwidth at most $(k-1,k)$
\end{itemize}

We say that a set of structures $C$ has pathwidth at most $(j,k)$ if every
structure $\aest$ in $C$ has pathwidth at most $(j,k)$. 

\end{definition}

\section{Pebble-Relation Games}

In this section we will introduce a game, called $(j,k)$-pebble-relation game, that captures expressibility in
$\pathlogic^{j,k}$.

Let $S_1$ and $S_2$ be two (not necessarily finite) sets. A {\em relation} $T$ with domain $S_1$ and range $S_2$ is
a collection of functions with domain $S_1$ and range $S_2$.
Remark: some confusion can arise from
the fact that generally (and in this paper) the name relation is used with another meaning; for example, 
an $r$-ary relation over $B$ is a subset of $B^r$. Both concepts are perfectly consistent, since
an $r$-ary relation over $B$ is, indeed, a relation in our sense with domain $\{1,\dots,r\}$ and
range $B$. 

Let $f$ be a function with domain $S_1$ and range $S_2$, and let $S_1'$ be a subset of its domain $S_1$. 
We will denote by $f_{|S'_1}$ the restriction of $f$ to $S_1$. Similarly, let
$T$ be a relation with domain $S_1$ and range $S_2$, and let $S'_1$ be a subset of its domain $S_1$.
We will denote by $T_{|S'_1}$ the relation with domain $S'_1$ and range $S_2$ that
contains $f_{|S'_1}$ for every $f\in T$. For every relation $T$ we denote by $\dom(T)$ the
domain of $T$.
We have two relations with domain $\emptyset$: the relation $\{\lambda\}$ 
and the relation $\emptyset$.

Let $0\leq j\leq k$ be non-negative integers and let $\aest$ and $\best$ be (not necessarily finite) $\vocab$-structures.
The $(j,k)$-pebble-relation ($(j,k)$-PR) game on $\aest$ and $\best$ is played between
two players, the {\em Spoiler} and the {\em Duplicator}. 
A configuration of the game consists of a relation $T$ with domain 
$I=\{a_1,\dots,a_{k'}\}\subseteq A$,
$k'\leq k$ and range $B$ such that every function $f$ in $T$ is a homomorphism from
$\aest_{|I}$ to $\best$.

Initially $I=\emptyset$
and $T$ contains the (unique) homomorphism
from $\aest_{|\emptyset}$ to $\best$, that is, $\lambda$. Each round of the game consists of a 
move from the Spoiler and a move from the Duplicator. Intuitively,
the Spoiler has control on the domain $I$ of $T$, which can be regarded as placing
some pebbles on the elements of $A$ that constitute $I$, whereas the Duplicator
decides the content of $T$ after the domain $I$ has been set by the Spoiler.
There are two types
of rounds: {\em shrinking} rounds and {\em blowing} rounds. 

Let $T^n$ be the 
configuration after the $n$-th round. The spoiler decides whether the following
round is a blowing or shrinking round.

\begin{itemize}
\item
If the $(n+1)$-th round is a shrinking round,
the Spoiler sets $I^{n+1}$ (the domain of $T^{n+1}$) to be a subset of the domain $I^n$ of $T^n$.
The Duplicator
responds by projecting every function in $T^n$ onto
the subdomain defined by $I^{n+1}$, that is, $T^{n+1}=T^n_{|I^{n+1}}$.
\item A blowing round only can be performed if $|I^n|\leq j$. In this case the 
Spoiler sets $I^{n+1}$ to be a superset of $I^n$ with $|I^{n+1}|\leq k$.
The duplicator responds by providing a $T^{n+1}$ with domain $I^{n+1}$
such that $T^{n+1}_{|I^n}\subseteq T^n$. That is, $T^{n+1}$ should contain
some extensions of functions in $T^n$ over the domain $I^{n+1}$ (recall
that any such extension must be a homomorphism from $\aest_{|I^{n+1}}$ to $\best$).
\end{itemize}
The Spoiler wins the game if the response of the Duplicator sets $T^{n+1}$ to $\emptyset$,
i.e., the Duplicator could not extend successfully any of the functions.
Otherwise, the game resumes.
The Duplicator wins the game if he has an strategy that allows him to continue
playing ``forever'', i.e., if the Spoiler can never win a round of the game.

\BEGINCOMMENT
{\tt Decide whether to give an intuitive explanation of the relation
of the winning strategy with the game}
\ENDCOMMENT

Now, we will present an algebraic characterization of the $(j,k)$-PR game. 

\begin{definition}
Let $0\leq j\leq k$ be non-negative integers and let $\aest$ and $\best$ be (not necessarily finite) 
$\vocab$-structures.
We say that {\em the Duplicator has a winning strategy for the $(j,k)$-pebble-relation game}
on $\aest$ and $\best$ if there is a nonempty family $\mathcal H$ of relations
such that:
\begin{itemize}
\item[(a)] every relation $T$ has range $B$ and domain $I$ for some $I\subseteq A$ with $|I|\leq k$.
\item[(b)] for every relation $T$ in $\mathcal H$ with domain $I$, 
$\emptyset\neq T$ and $T\subseteq\hom(\aest_{|I},\best)$
\item[(c)] $\mathcal H$ is closed under restrictions: for every $T$ in $\mathcal H$ with
domain $I$ and every $I'\subseteq I$, we have that $T_{|I'}\in\mathcal H$.
\item[(d)] $\mathcal H$ has the $(j,k)$-forth property: for every
relation $T$ in $\mathcal H$ with domain $I$ with $|I|\leq j$ and
every superset $I'$ of $I$ with $|I'|\leq k$, there exists some relation $T'$ in $\mathcal H$
with domain $I'$ such that $T'_{|I}\subseteq T$.
\end{itemize}
Furthermore, if $\mathcal H$ satisfies the following condition we say that $\mathcal H$
is a {\em strict} winning strategy for the $(j,k)$-pebble-relation game.
\begin{itemize}
\item[(d')] $\mathcal H$ has the strict $(j,k)$-forth property: for every
relation $T$ in $\mathcal H$ with domain $I$ with $|I|\leq j$ and
every superset $I'$ of $I$ with $|I'|\leq k$, the relation with domain $I'$ given by
$\{h\in\hom(\aest_{|I'},\best):h_{|I}\in T\}$ belongs to $\mathcal H$.
\end{itemize}
\end{definition}

The intuition behind the definition of a winning strategy is that every relation in
a winning strategy corresponds to a winning configuration for the Duplicator in the game.

\BEGINCOMMENT
\begin{proposition}
Let $\aest$ and $\best$ be $\vocab$-structures, let $0\leq j\leq k$ be non-negative integers, and
let $\mathcal F$ be a collection of winning strategies for the Duplicator on the 
$(j,k)$-pebble-relation game on $\aest$ and $\best$, then also the union $\bigcup\mathcal F$
is a winning strategy for the Duplicator. Hence, there is a largest winning strategy for the Duplicator
for the $(j,k)$-pebble game, namely the union of all winning strategies, which we will
denote ${\mathcal H}^{j,k}(\aest,\best)$.
\end{proposition}
\ENDCOMMENT

We need an auxiliary definition that is used a number of times in the proofs.
\begin{definition}
Let $\aest$ be a $\vocab$-structure, let $a_1,\dots,a_k$ be (not necessarily different)
elements of $A$, and let $v_1,\dots,v_k$ be variables. We denote by $\canonic(\aest,a_1,\dots,a_k)(v_1,\dots,v_k)$
the formula in $\conjrestpathlogic^{k,k}$, {\em with equality}, with variables among $v_1,\dots,v_k$ defined by
$$\bigwedge_{R\in\vocab}\bigwedge_{\langle a_{l_1},\dots,a_{l_{\arity(R)}}\rangle\in 
R^{\aest}} 
R(v_{l_1},\dots,v_{l_{\arity(R)}})\wedge\bigwedge_{1\leq i<j\leq k, a_i=a_j}(v_i=v_j)$$
\end{definition}

Notice that if all the elements $a_1,\dots,a_k$ are different then we do not need the equality.

The following properties of $\canonic(\aest,a_1,\dots,a_k)(v_1,\dots,v_k)$
will be very helpful.

\begin{proposition}
\label{pro:canonic}
  Let $\aest$ be a $\vocab$-structure, let $a_1,\dots,a_k$ be elements
  of $A$ and let $\theta$ be the formula
  $\canonic(\aest,a_1,\dots,a_k)(v_1,\dots,v_k)$. Then we have
\begin{itemize}
\item For every $\vocab$-structure $\best$ and every $b_1,\dots,b_k\in
B$, we have that
$$\best,b_1,\dots,b_k\models\theta(v_1,\dots,v_k) \text{ iff }
\aest_{|\{a_1,\dots,a_k\}},a_1,\dots,a_k\maps\best,b_1,\dots,b_k$$
\item For every quantifier-free formula $\varphi(v_1,\dots,v_k)$ in $L^k$
which statisfies %  such that 
$\aest,a_1,\dots,a_k\models\varphi(v_1,\dots,v_k)$ we have that $\theta(v_1,\dots,v_k)$
implies $\varphi(v_1,\dots,v_k)$, i.e., for every $\best$ and every $b_1,\dots,b_k\in B$,
$$\best,b_1,\dots,b_k\models (\theta\Rightarrow\varphi)(v_1,\dots,v_k)$$
\end{itemize}
\end{proposition}

The proof of these two simple facts about $\canonic(\aest,a_1,\dots,a_k)(v_1,\dots,v_k)$ follows the lines of the proof of a well-known result, due to Chandra
and Merlin~\cite{Chandra/Merlin:1977} which states that conjunctive query evaluation, conjunctive query 
containment, and deciding the existence of a homomorphism are essentially the same problem.

The following results show that pebble-relation games, expressiveness in the existential positive fragment of finite-variable infinitary logic with
restricted conjunction, and obstruction sets of bounded pathwidth are equivalent mathematical embodiments of the same concept.

\begin{theorem}
\label{the:gamesformulas}
Let $0\leq j\leq k$ be non-negative integers and let $\aest$ and $\best$ be (not necessarily finite) $\vocab$-structures.
The following statements are equivalent:
\begin{enumerate}
\item The Duplicator has a winning strategy $\mathcal H$ for
the $(j,k)$-PR game on $\aest$ and $\best$.
\item For every sentence $\varphi$ in $\pathlogic^{j,k}$ such that $\aest\models\varphi$ we
have that $\best\models\varphi$.
\item For every sentence $\varphi$ in $\restpathlogic^{j,k}$ such that $\aest\models\varphi$ we
have that $\best\models\varphi$.
\item For every sentence $\varphi$ in $\conjrestpathlogic^{j,k}$ such that $\aest\models\varphi$ we
have that $\best\models\varphi$.
\item Every $\vocab$-structure $\pest$ with pathwidth at most $(j,k)$ that homomorphically maps
to $\aest$ also homomorphically maps to $\best$.
\end{enumerate}
The equivalences hold even if we allow the use of equalities in the formulas.
\end{theorem}

\proof

\imply{1}{2} Let $\mathcal H$ be a winning strategy for the Duplicator. We shall show, by
induction on the construction of $\pathlogic^{j,k}$ formulas, that if $\varphi(v_1,\dots,v_m)$
is a formula of $\pathlogic^{j,k}$ {\em with equalities} where the free variables of $\varphi$ are among $v_1,\dots,v_m$, 
then the following property (*) holds:

(*) For any elements $a_1,\dots,a_m$ in $A$ such that 
$\aest,a_1,\dots,a_m\models\varphi(v_1,\dots,v_m)$, there exists some
$T\in\mathcal H$ with domain $I\subseteq\{a_1,\dots,a_m\}$ 
such that for every $h:\{a_1,\dots,a_m\}\rightarrow B$ such that $h_{|I}\in T$
we have
$$\best,h(a_1),\dots,h(a_m)\models\varphi(v_1,\dots,v_m).$$

First, it is easy to see that if $\varphi(v_1,\dots,v_m)$ is a quantifier-free formula in $\pathlogic^{j,k}$
such that $\aest,a_1,\dots,a_m\models\varphi(v_1,\dots,v_m)$ and
$h:\{a_1,\dots,a_m\}\rightarrow B$ is any homomorphism from $\aest_{|\{a_1,\dots,a_m\}}$ to $\best$ then 
$\best,h(a_1),\dots,h(a_m)\models\varphi(v_1,\dots,v_m)$. 

To see it, let
define $\theta$ as $\canonic(\aest,a_1,\dots,a_m)(v_1,\dots,v_m)$. By Proposition~\ref{pro:canonic} we have
$\best,h(a_1),\dots,h(a_m)\models\theta(v_1,\dots,v_m)\Rightarrow\varphi(v_1,\dots,v_m)$. Consequently, We can
infer that $\best,h(a_1),\dots,h(a_m)\models\varphi(v_1,\dots,v_m)$
from the fact 
that $\best,h(a_1),\dots,h(a_m)\models\theta(v_1,\dots,v_m)$.

Thus, any relation $T\in\mathcal H$
with domain $\{a_1,\dots,a_m\}$ would satisfy condition (*).
The inductive step for infinitary disjunction $\bigvee$ is straightforward using the induction
hypothesis.

Assume that the formula $\varphi(v_1,\dots,v_m)$ is of the form $\bigwedge\Psi$ where $\Psi$
is a collection of formulas in $\pathlogic^{j,k}$. We have to show that there exists some 
$T\in\mathcal H$ with domain $I\subseteq\{a_1,\dots,a_m\}$ such that for every
$h:\{a_1,\dots,a_m\}\rightarrow B$ with $h_{|I}\in T$
we have
$\best,h(a_1),\dots,h(a_m)\models\varphi(v_1,\dots,v_m)$.
Assume that there exists some formula
$\gamma$ in $\Psi$ which is quantified but not a sentence (such formula
if exists must be unique). Since $\gamma$ has at most $j$ variables we have
that $\gamma=\gamma(v_{i_1},\dots,v_{i_{j'}})$ where $j'\leq j$.
Thus, $\aest,a_{i_1},\dots,a_{i_{j'}}\models\gamma(v_{i_1},\dots,v_{i_j'})$, 
and by the induction hypothesis there exists some $T'\in\mathcal H$ with domain $I'\subseteq\{a_{i_1},\dots,a_{i_{j'}}\}$
such that for all $h:\{a_{i_1},\dots,a_{i_{j'}}\}\rightarrow B$ such that $h_{|I'}\in T'$, we
have $\best,h(a_{i_1}),\dots,h(a_{i_{j'}})\models\gamma(v_{i_1},\dots,v_{i_{j'}})$. If such formula does
not exists then set $I'=\emptyset$ and $T'=\{\lambda\}$.

By the $(j,k)$-forth property,
there exists some relation $T$ with domain $\{a_1,\dots,a_m\}$ such that $T_{|I'}\subseteq T'$. We shall see that $T$ satisfies
the desired condition. Let $h:\{a_1,\dots,a_m\}\rightarrow B$ such that $h\in T$,
and let $\psi(v_1,\dots,v_m)$ be any formula in $\Psi$. We have to study three cases:
\begin{itemize}
\item If $\psi(v_1,\dots,v_m)$ is quantifier-free then since 
$h\in\hom(\aest_{|\{a_1,\dots,a_m\}},\best)$, we have that $\best,h(a_1),\dots,h(a_m)\models\psi(v_1,\dots,v_m)$.
\item If $\psi(v_1,\dots,v_m)$ is a sentence then by induction hypothesis we have $\best\models\psi$ and thus, in consequence,
$\best,h(a_1),\dots,h(a_m)\models\psi(v_1,\dots,v_m)$.
\item Otherwise, we have $\psi(v_1,\dots,v_m)=\gamma(v_{i_1},\dots,v_{i_{j'}})$. Since $T_{|I'}\subseteq T'$,
we have $\best,h(a_{i_1}),\dots,h(a_{i_{j'}})\models\gamma(v_{i_1},\dots,v_{i_{j'}})$.
\end{itemize}

With respect to the existential quantification we shall distinguish two cases, depending on
whether or not the variable quantified is contained in $v_1,\dots,v_m$.

First, assume that the formula $\varphi(v_1,\dots,v_m)$ is of the form
$(\exists v)\psi(v_1,\dots,v_m,v)$ (i.e., $v\not\in\{v_1,\dots,v_m\}$). 
Thus, there exists some $a\in A$
such that $\aest,a_1,\dots,a_m,a\models\psi(v_1,\dots,v_m,v)$. Consequently,
by the induction hypothesis, there exists some $T'\in\mathcal H$ with
domain $I'\subseteq\{a_1,\dots,a_m,a\}$ such that for every 
$h:\{a_1,\dots,a_m,a\}\rightarrow B$ such that $h_{|I'}\in T'$
we have $\best,h(a_1),\dots,h(a_m),h(a)\models\psi(v_1,\dots,v_m,v)$.
We consider two cases:
\begin{itemize}
\item
If $a\in\{a_1,\dots,a_m\}$ then we set $I=I'$ and $T=T'$. Let 
$h:\{a_1,\dots,a_m\}\rightarrow B$ such that $h_{|I}\in T$. In consequence,
we have
$\best,h(a_1),\dots,h(a_m),h(a)\models\psi(v_1,\dots,v_m,v)$, and, in consequence, 
$\best,h(a_1),\dots,h(a_m)\models\psi(v_1,\dots,v_m)$.
\item Otherwise, we set $I=I'\backslash\{a\}$ and $T=T'_{|I}$ (notice
that if $a\not\in I'$ then $I=I'$ and $T=T'$). Let
$h:\{a_1,\dots,a_m\}\rightarrow B$ be a mapping such that $h_{|I}\in T$
and let $h':\{a_1,\dots,a_m,a\}$ be an extension of $h_{|I}$ such
that $h'_{|I'}\in T'$ (such extension always exists since
$T=T'_{|I}$). Thus
$\best,h'(a_1),\dots,h'(a_m),h'(a)\models\psi(v_1,\dots,v_m)$ and
consequently
$\best,h(a_1),\dots,h(a_m)\models\psi(v_1,\dots,v_m)$.
\end{itemize}

Secondly, assume that $\varphi(v_1,\dots,v_m)$ is
of the form $(\exists v)\psi(v_1,\dots,v_m)$. We can assume without
loss of generality that 
$\{v_1,\dots,v_{m'},v\}=\{v_1,\dots,v_m\}$ for $m'=m-1$.

We have that 
$\aest,a_1,\dots,a_{m'}\models\varphi(v_1,\dots,v_{m'})$.
Thus, we are in the previous case and, in consequence, there exists some relation $T$ 
with domain $I\subseteq\{a_1,\dots,a_{m'}\}$ such that 
for every $h':\{a_1,\dots,a_{m'}\}\rightarrow B$ with $h'_{|I}\in T$ we have 
$\best,h'(a_1),\dots,h'(a_{m'})\models\varphi(v_1,\dots,v_{m'})$. Finally, it is
easy to see that $T$ satisfies property (*):
Let $h:\{a_1,\dots,a_m\}\rightarrow B$ such that $h_{|I}\in T$
and let $h'=h_{|\{a_1,\dots,a_{m'}\}}$. Since $h_{|I}=h'_{|I}\in T$ we have
that $\best,h'(a_1),\dots,h'(a_{m'})\models\varphi(v_1,\dots,v_{m'})$ and
hence $\best,h(a_1),\dots,h(a_m)\models\varphi(v_1,\dots,v_m)$.

\imply{2}{3} and \imply{3}{4}. Straightforward since $\conjrestpathlogic^{j,k}\subseteq\restpathlogic^{j,k}\subseteq\pathlogic^{j,k}$.

\imply{4}{5}
It will be convenient to assume that the path decomposition has a sort of
canonical form. We will say that a path-decomposition $S_1,\dots,S_n$ of a structure $\pest$ is {\em canonical}
if: (a) $S_n=\emptyset$ and (b) for every $1\leq i\leq n-1$ we have that $S_i\subseteq S_{i+1}$
or $S_i\supseteq S_{i+1}$. It is easy to verify that if a structure $\pest$ has 
pathwidth at most $(j,k)$, $0\leq j\leq k$, then it has a canonical path-decomposition of width $(j,k)$.

Let $\pest$ be a structure and let $S_1,\dots,S_n$ be a canonical path-decomposition of width $(j,k)$
of $\pest$ where $S_1=\{p_1,\dots,p_m\}$, $m\leq k$. We shall show, by induction
on the size $n$ of the decomposition, that there exists a formula $\varphi(v_1,\dots,v_m)$
in $\conjrestpathlogic^{j,k}$ with variables among $v_1,\dots,v_k$ and whose free variables are exactly
$v_1,\dots,v_m$ such that for every 
structure $\dest$ and every $d_1,\dots,d_m\in D$ we have that
$$\pest,p_1,\dots,p_m\maps\dest,d_1,\dots,d_m,$$
if and only if
$$\dest,d_1,\dots,d_m\models\varphi(v_1,\dots,v_m).$$

The result then follows from the following line of reasoning: Let $\pest$ be any structure of pathwidth at most $(j,k)$ and
let, $S_1,\dots,S_m$ be the canonical path-decomposition that certifies its pathwidth. Then $\emptyset,S_1,\dots,S_m$
is also a path-decomposition of $\pest$ and, consequently, there exists a sentence $\varphi$ in $\conjrestpathlogic^{j,k}$
such that for every structure $\dest$, $\pest\maps\dest$ iff $\dest\models\varphi$. Thus, if $\pest\maps\aest$ then
$\aest\models\varphi$ and by condition (4), $\best\models\varphi$, and $\pest\maps\best$.

The base case in the induction ($n=0$) is easily proved by setting $\varphi$ to
the formula identically true which can be obtained as the conjunction of
an empty set of formulas.

For the induction step, let $S_1,\dots,S_n,S_{n+1}$ be a path-decomposition of $\pest$
of width $(j,k)$. Thus $S_2,\dots,S_n,S_{n+1}$ is a path-decomposition of 
$\pest_{|S_2\cup\dots\cup S_{n+1}}$. We should distinguish two cases, depending
on whether $S_1\subseteq S_2$ or $S_2\subseteq S_1$.

First assume that $S_1\subseteq S_2$. Assume, without loss of generality that
$S_1=\{p_1,\dots,p_m\}$ and $S_2=\{p_1,\dots,p_l\}$ where $0\leq m\leq l\leq k$ and $m\leq j$.
Notice that since $S_1\subseteq S_2$, $S_2,\dots,S_{n+1}$ is a path-decomposition
of $\pest$. Thus, by the inductive hypothesis, there exists some formula $\varphi(v_1,\dots,v_l)$
in $\conjrestpathlogic^{j,k}$, 
such that for every $\vocab$-structure $\dest$ and every $d_1,\dots,d_l\in D$
$$\pest,p_1,\dots,p_l\maps\dest,d_1,\dots,d_l$$
if and only if 
$$\dest,d_1,\dots,d_l\models\varphi(v_1,\dots,v_l)$$ 

Consider the formula $\psi(v_1,\dots,v_m)$ in $\conjrestpathlogic^{j,k}$ with variables
among $v_1,\dots,v_k$ and whose free variables are exactly $v_1,\dots,v_m$ defined
by:
$$\psi(v_1,\dots,v_m)=(\exists v_{m+1},\dots,v_l)\varphi(v_1,\dots,v_l)$$
We shall show that $\psi$ satisfies the desired property. For every 
$\dest$ and every $d_1,\dots,d_m\in D$ we have that
$$\pest,p_1,\dots,p_m\maps\dest,d_1,\dots,d_m,$$
if and only if 
$$\exists d_{m+1},\dots,d_l\in D \text{ s.t. } \pest,p_1,\dots,p_l\maps\dest,d_1,\dots,d_l,$$
By the by the induction hypothesis is equivalent to 
$$\exists d_{m+1},\dots,d_l\in D \text{ s.t.} \dest,d_1,\dots,d_l\models\varphi(v_1,\dots,v_l),$$
which is equivalent to
$$\dest,d_1,\dots,d_m\models(\exists v_{m+1},\dots,v_l)\varphi(v_1,\dots,v_l)=\psi(v_1,\dots,v_m)$$

Assume now that $S_2\subseteq S_1$. Let $S_1=\{p_1,\dots,p_m\}$ and $S_2=\{p_1,\dots,p_l\}$, 
where $0\leq l\leq m\leq k$ and $l\leq j$.
By the inductive hypothesis there exists some formula
$\varphi(v_1,\dots,v_l)$ in $\conjrestpathlogic^{j,k}$, with variables among
$v_1,\dots,v_k$ and whose free variables are exactly
$v_1,\dots,v_l$ such that for every $\vocab$-structure $\dest$ and every $d_1,\dots,d_l\in D$
$$\pest_{|S_2\cup\dots\cup S_{n+1}},p_1,\dots,p_l\maps\dest,d_1,\dots,d_l$$
if and only if
$$\dest,d_1,\dots,d_l\models\varphi(v_1,\dots,v_l).$$

Let $\theta(v_1,\dots,v_m)\in\conjrestpathlogic^{j,k}$
defined by $\theta=\canonic(\pest,p_1,\dots,p_m)(v_1,\dots,v_m)$. Notice that
since $p_1,\dots,p_m$ are different, $\theta$ does not contain equality.
Finally consider the formula $\psi(v_1,\dots,v_m)$ in $\conjrestpathlogic^{j,k}$ defined by:
$$\psi(v_1,\dots,v_m)=\theta(v_1,\dots,v_m)\wedge\varphi(v_1,\dots,v_l)$$

We shall prove that $\psi$ satisfies the desired property. Let $\dest$ be a $\vocab$-structure
and let $d_1,\dots,d_m$ be elements in $D$. Let $h$ be any homomorphism from $\pest$ to $\dest$
such that $h(p_i)=d_i$ for $1\leq i\leq m$. 

Thus $h_{|S_1}$ is
a homomorphism from $\pest_{|S_1}$ to $\dest$ and, by Proposition~\ref{pro:canonic},
$\dest,d_1,\dots,d_m\models\theta(v_1,\dots,v_m)$.
Also, we have that $h_{|S_2\cup\dots\cup S_{n+1}}$ is a homomorphism from 
$\pest_{|S_2\cup\dots\cup S_{n+1}}$ to $\dest$ and by the induction hypothesis,
$\dest,d_1,\dots,d_l\models\varphi(v_1,\dots,v_l)$. Putting all together we have
that 
$$\dest,d_1,\dots,d_m\models\theta(v_1,\dots,v_m)\wedge\varphi(v_1,\dots,v_l)=\psi(v_1,\dots,v_m)$$

Conversely, suppose that $\dest,d_1,\dots,d_m\models\psi(v_1,\dots,v_m)$. In consequence we have that
$\dest,d_1,\dots,d_l\models\varphi(v_1,\dots,v_l)$ and, by induction hypothesis,
there exists some homomorphism $h'$ from $\pest_{|S_2\cup\dots\cup S_{n+1}}$ to
$\dest$ such that $h'(p_i)=d_i$, $1\leq i\leq l$. Let $h:P\rightarrow D$ be the mapping defined by
$$h(p)=\left\{\begin{array}{ll}
d_i & \text{ if } p=p_i \\
h'(p) & \text{ otherwise } \\
\end{array}\right.
$$
We shall show that $h$ defines a homomorphism from $\pest$ to $\dest$.
First notice that since $\dest,d_1,\dots,d_m\models\theta(v_1,\dots,v_m)$, 
$h_{|S_1}$ defines a homomorphism from $\pest_{|S_1}$ to $\dest$. 
Finally, notice that $h_{|S_2\cup\dots\cup S_{n+1}}=h'$
defines a homomorphism from $\pest_{|S_2\cup\dots\cup S_{n+1}}$ to $\dest$.
Then we are at home: Let $R$ be any relation symbol in $\vocab$ of arity, say $r$,
and let $\langle a_1,\dots,a_r\rangle$ any tuple in $R^{\pest}$. Since $S_1,\dots,S_{n+1}$
is a path-decomposition of $\pest$ we have that $\{a_1,\dots,a_r\}\in S_j$ for some $j\in\{1,\dots,n+1\}$.
If $j=1$ then we have that $\langle h(a_1),\dots,h(a_r)\rangle$ in $R^{\best}$, as $h_{|S_1}$ defines
a homomorphism from $\pest_{|S_1}$ to $\dest$. If, otherwise, $j>1$, then $\langle h(a_1),\dots,h(a_r)\rangle$
is in $R^{\dest}$, as $h_{|S_2\cup\cdots\cup S_{n+1}}$ defines a homomorphism from $\pest_{|S_2\cup\cdots\cup S_{n+1}}$

\imply{5}{1} We shall produce a winning strategy $\mathcal H$ for the Duplicator. For every
structure $\pest$, for every
path-decomposition $S_1,\dots,S_n$ of width $(j,k)$ of $\pest$, and for every mapping $h:P\rightarrow A$ such
that:
\begin{itemize}
\item[(i)] $h$ is a homomorphism from $\pest$ to $\aest$ and
\item[(ii)] $h_{|S_1}$ is one-to-one and furthermore $(h_{|S_1})^{-1}$
is a homomorphism from $\aest_{|h(S_1)}$ to $\pest_{S_1}$.
\end{itemize}

The set $\mathcal H$ contains the relation
$T$ with domain $h(S_1)$ defined by
$$T=\{g_{|S_1}\comp (h_{|S_1})^{-1}:\pest\stackrel{g}\maps\best\}$$

First, notice that if $\pest$ is the structure with universe $P=\emptyset$, and $h$ and $g$ are $\lambda$, then
we have that $\{\lambda\}\in {\mathcal H}$, and hence, $\mathcal H$ is non-empty. 
We show now that $\mathcal H$ has the required properties:
\begin{itemize}
\item Clearly, every relation $T$ in $\mathcal H$ has domain $I\subseteq A$
with $|I|\leq k$ and range $B$.
\item We have to show that for every $T$ with domain $I$ and every $f\in T$, $f$ is a homomorphism from $\aest_{|I}$ 
to $\best$. First, assume that $f=g_{|S_1}\comp h_{|h(S_1)}^{-1}$ where $g$, $h$ and $S_1$ are as defined above. Since
$h_{|S_1}^{-1}$ is a homomorphism from $\aest_{|h(S_1)}$ to $\pest_{|S_1}$ and
$g_{|S_1}$ is a homomorphism from $\pest_{|S_1}$ to $\best$, then its composition $f$
must be a homomorphism from $\aest_{|h(S_1)}$ to $\best$. Furthermore, $T$ is non empty, since
$\pest\maps\aest$ implies that $\pest\maps\best$.
\item We have to show that $\mathcal H$ is closed under projection. Let $T$
be any relation in $\mathcal H$ obtained from $\pest$, $S_1,\dots,S_n$ and $h$
as defined above. In consequence $T$ has domain $h(S_1)$. Let $I\subseteq h(S_1)$ and
let $S'_1=h^{-1}(I)$.
It is not difficult to
see that $S'_1,S_1,\dots,S_n$ defines a path 
decomposition of $\pest$, that $h_{|S'_1}$ is one-to-one, and that $(h_{|S'_1})^{-1}$
defines a homomorphism from $\aest_{|I}$ to $\pest_{|S'_1}$. Thus the relation
$T'=\{g_{|S'_1}\comp (h_{|S'_1})^{-1}:\pest_{|S'_1}\stackrel{g}\maps\best\}$ with domain $I$
belongs also to $\mathcal H$. We shall show that $T_{|I}=T'$. For every homomorphism $f'$
from $\aest_{|I}$ to $\best$, $f'\in T'$ iff there exists some homomorphism $g$ from
$\pest$ to $\best$ such that $f'=g_{|S'_1}\comp h_{|S'_1}$. In consequence, $f'\in T'$ iff
$f=g_{|S_1}\comp h_{S_1}\in T$. Finally, the result follows from the fact that $f_{|S'_1}=f'$.

\item We have to show that $\mathcal H$ has the $(j,k)$-forth property. Let
$T$ be any relation in $\mathcal H$ with domain $I$ obtained from $\pest$, 
$S_1,\dots,S_n$ and $h$ as defined above and let $I'$ be any superset of $I$
with $|I'|\leq k$. Let $I'\backslash I=\{a'_1,\dots,a'_l\}$ and let $p'_1,\dots,p'_l$
elements not in $P$. Let $P'=P\cup\{p'_1,\dots,p'_l\}$, let $S'_1=S_1\cup\{p'_1,\dots,p'_l\}$
and let $h':P'\rightarrow A$ be the extension of $h$ that maps $p'_i$ to $a'_i$ $(1\leq i\leq l)$. That is
$$h'(p)=\left\{\begin{array}{ll}
a'_i & \text{if } p=p'_i, 1\leq i\leq l \\
h(p)& \text{otherwise}
\end{array}
\right.
$$
We define the structure $\pest'$ as the structure
with universe $P'$ such that for every $R$ in $\vocab$, 
$$R^{\pest'}=
R^{\pest}\cup\{\langle p_1,\dots,p_m\rangle: \langle h'(p_1),\dots,h'(p_m)\rangle\in R^{\aest}, 
\{p_1,\dots,p_m\}\subseteq S'_1\}$$
Notice that by construction $S'_1,S_1,\dots,S_n$ is a path-decomposition
of width $(j,k)$ of $\pest'$ and that $h'$ is a homomorphism from $\pest'$ to $\aest$ such
that $h'_{|S'_1}$ is one-to-one and that $(h'_{|S'_1})^{-1}$
is a homomorphism from $\aest_{|I'}$ to $\pest_{|S'_1}$.
Let $T'$ be the relation obtained from $\pest'$, $S'_1,S_1,\dots,S_n$ and $h'$.
We shall see that $T'_{|I}\subseteq T$: Let $f$ be any function in $T'$. Thus,
$f=g_{S'_1}\comp (h'_{|S'_1})^{-1}$ for some homomorphism $g$ from $\pest'$
to $\best$. Thus, $g_{|S_1\cup\dots\cup S_n}$ defines a homomorphism from $\pest$ to $\best$
and, in consequence, $f_{|I}=g_{|S_1}\comp (h_{|S_1})^{-1}$ belongs to $T$. 
\end{itemize} 

\endproof

The following theorem provides us with several alternative ways to characterize $\pathlogic^{j,k}$
definability.

\begin{theorem}
\label{the:gamesexpressivity}
Let $0\leq j\leq k$ be non-negative integers and 
let $\mathcal C$ be a class of $\vocab$-structures.
The following statements are equivalent:
\begin{enumerate}
\item The class $\mathcal C$ is $\pathlogic^{j,k}$-definable, i.e., there is a sentence
$\varphi$ of $\pathlogic^{j,k}$ such that for every $\vocab$-structure we have that
$\aest\in{\mathcal C}$ iff $\aest\models\varphi$. 
\item If $\aest$ and $\best$ are  $\vocab$-structures such that $\aest\in\mathcal C$
and the Duplicator has a winning strategy for the $(j,k)$-PR game on $\aest$ and $\best$, then 
$\best\in\mathcal C$.
\end{enumerate}
Furthermore, if $\neg\mathcal C$ is a finitely generated ideal we also have that $(1)$ and $(2)$
are equivalent to the following:
\begin{enumerate}
\item[(3)] The class $\mathcal C$ is $\restpathlogic^{j,k}$-definable.
\item[(4)] The class $\neg\mathcal C$ has an obstruction set of pathwidth at most $(j,k)$
\end{enumerate}

\end{theorem}
\proof

First, we will need the following definition.

Let $\mathcal S$ be a finite collection of $\vocab$-structures. We say that
$\mathcal S$ is disjoint if for every
different $\aest,\best\in\mathcal S$, $A\cap B=\emptyset$. If $\mathcal S$ is
disjoint we define $\oplus\mathcal S$ as the  $\vocab$-structure whose
universe is the union of the universes of all the structures in $\mathcal S$, 
and such that for every relation symbol $R\in\vocab$,
$R^{\oplus\mathcal S}=\bigcup_{\aest\in\mathcal S} R^{\aest}$.
Observe that, as the universes of the structures in $\mathcal S$ are disjoint, if $\mathcal S$
has pathwidth at most $(j,k)$ for some $j,k$ then $\oplus\mathcal S$ has also pathwidth at most $(j,k)$.

If $\mathcal S=\{\aest_1,\dots,\aest_m\}$ we also denote $\oplus\mathcal S$
by $\aest_1\oplus\dots\oplus\aest_m$.

Let $\best$ be a $\vocab$-structure and let $\mathcal S$ be a set of $\vocab$-structures.
It is easy to verify that $\oplus{\mathcal S}\maps\best$ iff $\aest\maps\best$ for every $\aest\in\mathcal S$.

Now we are in a position to prove Theorem~\ref{the:gamesexpressivity}

\imply{1}{2} Straightforward from Theorem~\ref{the:gamesformulas}. Assume that 
$\aest\models\varphi$. Since the Duplicator 
has a winning strategy for the $(j,k)$-PR game then $\best\models\varphi$ and thus
$\best$ belongs to $\mathcal C$.

\imply{2}{1} Clearly $\mathcal C$ is a filter and
let $\neg\mathcal C$ be $I(\mathcal S)$ for some set $\mathcal S$ of $\vocab$-structures.
For every $\best$ in $\mathcal S$, let $\Psi_{\best}$ 
be the collection of sentences in $\restpathlogic^{j,k}$ falsified by $\best$ and 
let $\varphi_{\best}=\bigvee\Psi_{\best}$,
which is a sentence of $\restpathlogic^{j,k}$. Let $\Psi$ be the set containing
$\varphi_{\best}$ for every $\best$ in $\mathcal S$, and let $\varphi$ be 
the sentence of $\pathlogic^{j,k}$ defined by $\varphi=\bigwedge\Psi$. Notice that if
$\mathcal S$ is finite then $\varphi$ is in $\restpathlogic^{j,k}$. We shall see
that $\varphi$ defines $\mathcal C$. Let $\aest$ be any structure not in
$\mathcal C$. Thus, there exists some $\best$ in $\mathcal S$ such 
that $\aest$ homomorphically maps to $\best$. Recall that if $\aest$ is
homomorphic to $\best$, by Proposition~\ref{pro:hompressat}, $\best$
satisfies all sentences in $\pathlogic^{\omega}$ satisfied by $\aest$. 
In consequence, 
$\aest\not\models\varphi_{\best}$ and thus $\aest\not\models\varphi$.
Conversely, let $\aest$ be any structure such that
$\aest\not\models\varphi$, thus $\aest\not\models\varphi_{\best}$ for
some $\best$ in $\mathcal S$. Thus, every formula $\psi$ in
$\restpathlogic^{j,k}$ satisfied by $\aest$ is also satisfied by $\best$,
and, in consequence, the Duplicator has a winning strategy for
the $(j,k)$-pebble-relation game on $\aest$ and $\best$. Since $\best\not\in\mathcal C$
we conclude that $\aest\not\in C$.

\imply{3}{1} Trivial as $N^{j,k}\subseteq M^{j,k}$.

\imply{4}{1} Straightforward from Theorem~\ref{the:gamesformulas}.
Let $\aest\in\mathcal C$ and 
$\best\not\in\mathcal C$ be $\vocab$-structures. Since $\aest\in\mathcal C$ there exists
a structure $\pest$ in the obstruction set (and therefore of pathwidth at most $(j,k)$) 
that homomorphically maps to $\aest$. Since $\best\not\in\mathcal C$, $\pest$ does not
homomorphically map to $\best$. In consequence, there does not exist a winning strategy
for the Duplicator for the $(j,k)$-pebble-relation game on $\aest$ and $\best$.

\imply{2}{4} Let $\best_1,\dots,\best_m$ be the finite set of structures that
generates $\neg\mathcal C$.
Consider the set $\mathcal O$ given by the collection of all structures of the
form $\pest_1\oplus\dots\oplus\pest_m$ where $\pest_i$ is a $\vocab$-structure
of pathwidth at most $(j,k)$ that does not map to $\best_i$, and $\pest_i$, $1\leq i\leq m$
have disjoint universes. Thus, every structure in
$\mathcal O$ has pathwidth at most $(j,k)$.
We shall show that
$\mathcal O$ is an obstruction set of $\neg\mathcal C$.
First, observe that every structure $\pest_1\oplus\cdots\oplus\pest_m$ in $\mathcal O$
belongs to $\mathcal C$, as for each $i\in\{1,\dots,m\}$, $\pest_i\not\maps \best_i$.
Consequently, if one of such structures $\pest_1\oplus\cdots\oplus\pest_m$ is homomorphic
to a given structure $\aest$, then $\aest$ must be in $\mathcal C$, as $\mathcal C$ is
a filter.
Conversely, let $\aest$ be any structure in $\mathcal C$. Thus, for every
$1\leq i\leq m$, the Duplicator does not have a winning strategy for the
$(j,k)$-pebble game on $\aest$ and $\best_i$. Thus, for every $1\leq i\leq m$
there exists some $\vocab$-structure $\pest_i$ of pathwidth at most $(j,k)$ 
such that $\pest_i\maps\aest$ and $\pest_i\not\maps\best_i$ (we can assume
without loss of generality that the universes of $\pest_i$, $1\leq i\leq m$ 
are disjoint). Thus
$\pest_1\oplus\dots\oplus\pest_m$ belongs to $\mathcal O$ and
furthermore it homomorphically maps to $\aest$.

\imply{2}{3} The structure of this proof is similar to the previous case.

Let $\best_1,\dots,\best_m$ be the finite set of structures that
generates $\neg\mathcal C$.
Consider the set $\Psi$ given by the collection of all structures of the
form $\psi_1\wedge\dots\wedge\psi_m$ where $\psi_i$ is a sentence
in $O^{j,k}$ that is not satisfied by $\best_i$. Then the formula 
$\varphi$ defined to be $\bigvee \Psi$ is in $N^{j,k}$. Let us show
that $\varphi$ defines $\mathcal C$. First observe that for every
$\best_i$, $1\leq i\leq m$ generating $\neg\mathcal C$, we
have that $\best_i\not\models\varphi$. Consequently, as $\neg\mathcal C$ is an ideal, 
by Proposition~\ref{pro:hompressat}, for any structure $\aest$ not in $\mathcal C$,
$\aest\not\models\varphi$.
Conversely, let $\aest$ be any structure in $\mathcal C$. Thus, for every
$1\leq i\leq m$, the Duplicator does not have a winning strategy for the
$(j,k)$-pebble game on $\aest$ and $\best_i$. Thus, for every $1\leq i\leq m$
there exists some sentence $\psi_i$ in $O^{j,k}$ that is 
true on $\aest$ but not true on $\best_i$.
Thus $\psi_1\wedge\dots\wedge\psi_m$ is true on $\aest$ and belongs to $\Psi$.
Consequently, $\varphi$ is true on $\aest$.
\endproof

\section{Datalog Programs}

Let $\vocab$ be a vocabulary consisting of relational symbols. The class 
{\em SNP}~\cite{Kolaitis/Vardi:1987,Papadimitriou/Yannakakis:1991} is the
set of all existential second-order sentences with a universal first-order part, i.e.,
sentences of the form $\exists S_1,\dots,S_l\forall v_1,\dots,v_m\varphi(v_1,\dots,v_m)$
where $\varphi$ is a quantifier-free first-order formula over the
vocabulary $\vocab\cup\{S_1,\dots,S_l\}$ with variables among $v_1,\dots,v_m$.
We will assume that $\varphi$ is in CNF. We consider some restrictions that
can be enforced on the class SNP.
For {\em monotone SNP}~\cite{Feder/Vardi:1998}, we require every occurrence of a 
relation symbol from $\vocab$ to have negative polarity, i.e., a negation applied to it. 
For {\em $j$-adic SNP}, we require every second-order variable $S_i, 1\leq i\leq l$ to
have arity at most $j$. For $k$-ary, we require the number
of variables universally quantified $m$ to be at most $k$. 
For {\em Krom SNP} we require every clause of the quantifier-free first-order part
$\varphi$ to have at most two occurrences of a second-order variable.

Every Krom SNP formula $\varphi$ over the vocabulary $\vocab$ defines a class
of $\vocab$-structures, namely, the set containing every $\vocab$-structure $\aest$
such that $\aest\models\varphi$. Furthermore, the problem of deciding, given
a $\vocab$-structure $\aest$, whether $\aest\models\varphi$ is solvable in NL~\cite{Gradel:1992}.
Thus expressibility in Krom SNP is a sufficient condition for membership in NL.

For {\em restricted Krom SNP} we additionally require every clause of the quantifier-free first-order part
$\varphi$ to have at most one positive occurrence of a second-order variable and
at most one negative occurrence of a second-order variable.

\begin{theorem}
\label{the:expressivitygame}
Let $0\leq j\leq k$ be non-negative integers and let $\best$ be a $\vocab$-structure. 
There exists a sentence $\varphi$ in $j$-adic $k$-ary restricted Krom monotone SNP with equalities
such that for every $\vocab$-structure $\aest$, $\aest\models\varphi$ iff the Duplicator has
a winning strategy for the $(j,k)$-PR game on $\aest$ and $\best$.
\end{theorem}

\proof

The proof of Theorem~\ref{the:expressivitygame} requires some intermediate results. In a first step
we define a different notion of winning strategy, called ``supercomplete winning strategy'' and we
show that this new notion is equivalent to the notion of winning strategy introduced initially,
that is, we prove that for every $\vocab$-structures $\aest$ and $\best$, and for every $0\leq j\leq k$
the Duplicator has a winning strategy for the $(j,k)$-pebble game if and only if it
has a supercomplete winning strategy. In order to prove this we define several different but equivalent
notions of winning strategy that will act a links in a chain of inferences.

The new notions of winning strategy are the following:

\begin{definition}
Let $0\leq j\leq k$ be non-negative integers and let $\aest$ and $\best$ be $\vocab$-structures.
We say that the Duplicator has a {\em complete winning strategy} for the $(j,k)$-pebble-relation game
on $\aest$ and $\best$ if there is a  family $\mathcal H$ of relations such that:
\begin{itemize}
\item[(a)] every relation $T$ has range $B$ and domain $I$ for some $I\subseteq A$ with $|I|\leq j$.
\item[(b)] $\mathcal H$ contains $\{\lambda\}$ and does not contain $\emptyset$.
\item[(c)] for every $I\subseteq A$ with $|I|\leq k$, every relation $T$ in $\mathcal H$ 
with domain $I'\subseteq I$
and every $I''\subseteq I$ with $|I''|\leq j$, the relation with domain $I''$ given by
$$\{h_{|I''}: h\in\hom(\aest_{|I},\best), h_{|I'}\in T\}$$ 
belongs to $\mathcal H$
\end{itemize}
Furthermore, if $\mathcal H$ satisfies the following condition we say that $\mathcal H$ is a
{\em supercomplete winning strategy} for the $(j,k)$-pebble game
\begin{itemize}
\item[(d)] For every $T$ in $\mathcal H$ and every $T\subseteq T'$ we have that $T'\in\mathcal H$
\end{itemize}
\end{definition}

The following claim states that all the different characterizations of winning strategy introduced
so far are equivalent.

\begin{claim}
\label{cl:claim1}
Let $0\leq j\leq k$ be non-negative integers and let $\aest$ and $\best$ be $\vocab$-structures.
The following statements are equivalent:
\begin{enumerate}
\item The Duplicator has a winning strategy for the $(j,k)$-PR game on $\aest$ and $\best$.
\item The Duplicator has a strict winning strategy for the $(j,k)$-PR game on $\aest$ and $\best$.
\item The Duplicator has a complete winning strategy for the $(j,k)$-PR game on $\aest$ and $\best$.
\item The Duplicator has a supercomplete winning strategy for the $(j,k)$-PR game on $\aest$ and $\best$.
\end{enumerate}
\end{claim}
\proof

\imply{1}{2}
Let $\mathcal H$ be a winning strategy for the Duplicator for the $(j,k)$-PR game
on $\aest$ and $\best$. We define $\mathcal H^*$ to be the set
$$\{T^*: \dom(T^*)\subseteq A, |\dom(T^*)|\leq k, T^*\subseteq\hom(\aest_{\dom(T)},\best), \exists T\in{\mathcal H} 
\text{ s.t. } T\subseteq T^*\}$$
We will show that $\mathcal H^*$ is a strict winning strategy:
\begin{itemize}
\item It is obvious that $\mathcal H^*$ satisfies conditions (a) and (b).
\item $\mathcal H^*$ is closed under restrictions: Let $T^*$ a relation in $\mathcal H^*$ with domain $I$ 
and let $I'\subseteq I$. Thus, there exists some $T$ in $\mathcal H$ with domain $I$ such that $T\subseteq T^*$.
Since $\mathcal H$ is closed under restrictions $T_{|I'}\in\mathcal H$. Since $T_{|I'}\subseteq T^*_{|I'}$ we
have that $T^*_{|I'}\in\mathcal H^*$.
\item $\mathcal H^*$ has the strict $(j,k)$-forth property: Let $T^*$ be a relation in $\mathcal H^*$ with
domain $I$, with $|I|\leq j$ and let $I'$ be superset of $I$ with $|I'|\leq k$. Thus, there 
exists some relation $T$ in $\mathcal H$ with domain $I$ such that $T\subseteq T^*$. Since $\mathcal H$
has the $(j,k)$-forth property there exists some relation $T'$ in $\mathcal H$ with domain $I'$ such
that $T'_{|I}\subseteq T$. In consequence $T'\subseteq\{h\in\hom(\aest_{|I'},\best):h_{|I}\in T^*\}$.
Thus, the latter is in $\mathcal H^*$ 
\end{itemize}

\imply{2}{3} Let $\mathcal H$ be a strict winning strategy for the Duplicator and let $\mathcal H^*$
be the collection of relations in $\mathcal H$ with domain of size at most $j$. We shall show that $\mathcal H^*$
is a complete winning strategy. It is straightforward to show that $\mathcal H^*$ satisfies conditions 
(a) and (b) of the definition of a complete winning strategy. For condition (c), let $I\subseteq A$ with
$|I|\leq k$, let $T^*$ be a relation in $\mathcal H^*$ (and hence in $\mathcal H$)
with domain $I'\subseteq I$ and let $I''\subseteq I$
with $|I''|\leq j$. By the strict $(j,k)$-forth
property we have that the relation of arity $I$ given by $\{h\in\hom(\aest_{|I},\best),h_{|I'}\in T^*\}$
belongs also to $\mathcal H$ and so its projection to $I''$ which is given by
$\{h_{|I''}:h\in\hom(\aest_{|I},\best),h_{|I'}\in T^*\}$. Since $|I''|\leq j$, then it also belongs to $\mathcal H^*$.

\imply{3}{4}
Let $\mathcal H$ be a complete winning strategy and let $\mathcal H^*$ be the set containing every relation $T^*$
such that there exists a relation $T\in\mathcal H$ with $T\subseteq T^*$. Clearly $\mathcal H^*$ satisfies
conditions (a), (b), and (d) of a supercomplete winning strategy. For condition (c), let $I\subseteq A$ with
$|I|\leq k$, let $T^*$ be a relation in $\mathcal H^*$ with domain $I'\subseteq I$ and let $I''\subseteq I$
with $|I''|\leq j$. There exists some $T\in \mathcal H$ such that $T\subseteq T^*$ and, in consequence
$\{h_{|I''}:h\in\hom(\aest_{|I},\best),h_{|I'}\in T\}$ is a subset of $\{h_{|I''}:h\in\hom(\aest_{|I},\best),h_{|I'}\in T^*\}$
and hence the latter is in $\mathcal H^*$.

\imply{4}{1} 
Let $\mathcal H$ be a supercomplete winning strategy and let $\mathcal H^*$ be a set containing for every $I\subseteq A$
with $|I|\leq k$, and for every $T$ in $\mathcal H$ with domain $I'\subseteq I$, and for every $I''\subseteq I$
the relation
$\{h_{|I''}:h\in\hom(\aest_{|I},\best),h_{|I'}\in T\}$.
It is straightforward to show that $\mathcal H^*$ satisfies the conditions (a), (b), and (c) of a winning strategy. 
For condition (d), let $T''$ be any relation in $\mathcal H^*$ with domain $I''$ with $|I''|\leq j$. Thus, 
$T''=\{h_{|I''}:h\in\hom(\aest_{|I},\best),h_{|I'}\in T\}$ for some $T$ in $\mathcal H$ with domain $I'$ and
some $I'\subseteq I$ with $|I|\leq k$. Then, $T''$ is also in $\mathcal H$. Now, let $I'''$
be any superset of $I''$ with $|I'''|\leq k$. Thus, $T'''=\{h_{|I'''}:h\in\hom(\aest_{|I'''},\best),h_{|I''}\in T''\}$
is also in $\mathcal H$ and $T'''_{|I''}\subseteq T''$.

\endproof

In a second step we reformulate having a supercomplete winning strategy as the existence of a structure
satisfying certain properties.
Let $r$ be a non-negative integer and let $B$ a set. Recall that an $r$-ary relation over $B$ is a relation
with domain $\{1,\dots,r\}$ and range $B$.

Let $\{U_n:1\leq n\leq m\}$ be the collection of all $j$-ary relations over $B$ and let
$\{R_{U_n}:1\leq n\leq m\}$ be a collection of $j$-ary relation symbols, one for each relation $U_n:1\leq n\leq m$.
We shall show that the Duplicator has a superstrict winning strategy for the $(j,k)$-pebble-relation
game on $\aest$ and $\best$ iff there exists some structure $\aest'$ with domain $A$ over the
vocabulary $\{R_{U_n}:1\leq n\leq m\}$ such that:
\begin{itemize}
\item[(a)] $(R_{\emptyset})^{\aest'}=\emptyset$.
\item[(b)] For every (not necessarily different) $a_1,\dots,a_k\in A$, and for every 
$1\leq l_1,\dots,l_{j}\leq k$, $\langle a_{l_1},\dots,a_{l_{j}}\rangle\in {(R_U)}^{\aest'}$ where
$$U=\{\langle h(a_{l_1}),\dots,h(a_{l_{j}})\rangle:
h\in\hom(\aest_{\{a_1,\dots,a_k\}},\best)\}$$  
\item[(c)] For every (not necessarily different) $a_1,\dots,a_k$ elements in $A$, for every relation
symbol $R_{U_n}, 1\leq n\leq m$, 
for every $\langle a_{i_1},\dots,a_{i_{j}}\rangle\in (R_{U_n})^{\aest'}$ and for every 
$1\leq l_1,\dots,l_{j}\leq k$, we have that $\langle a_{l_1},\dots,a_{l_{j}}\rangle\in {(R_U)}^{\aest'}$ where
$$U=\{\langle h(a_{l_1}),\dots,h(a_{l_{j}})\rangle:
h\in\hom(\aest_{\{a_1,\dots,a_k\}},\best),
\langle h(a_{i_1}),\dots,h(a_{i_j})\rangle\in U_n\}$$
\item[(d)] For every $1\leq n,n'\leq k$ such that $U_n\subseteq U_{n'}$ we have 
$(R_{U_n})^{\aest'}\subseteq(R_{U_{n'}})^{\aest'}$ 
\end{itemize}

We shall say that $\aest'$ is a {\em relational} winning strategy for the Duplicator for 
the $(j,k)$-pebble-relation game on $\aest$ and $\best$.

\begin{claim}
\label{cl:claim2}
Let $0\leq j\leq k$ be non-negative integers and let $\aest$ and $\best$ be $\vocab$-structures.
The following statements are equivalent:
\begin{itemize}
\item The Duplicator has a supercomplete winning strategy for the $(j,k)$-pebble relation game on
$\aest$ and $\best$.
\item The Duplicator has a relational winning strategy for the $(j,k)$-pebble relation game on 
$\aest$ and $\best$.
\end{itemize}
\end{claim}
\proof
First, we need the following definition, let $a_1,\dots,a_l$ be a
collection of (not necessarily different) elements of $A$ and let $T$ be 
a relation with domain $\{a_1,\dots,a_l\}$. 
We define $U(T,a_1,\dots,a_l)$ be the $l$-ary relation over $B$ given by
$$\{\langle h(a_1),\dots,h(a_l)\rangle : h\in T \}$$

Let $\mathcal H$ be a supercomplete winning strategy for the Duplicator for the $(j,k)$-pebble
on $\aest$ and $\best$ and let $\aest'$ be the $\{R_{U_n}:1\leq n\leq m\}$-structure such that
for every $1\leq n\leq m$, 
$$
\begin{array}{ll}
(R_{U_n})^{\aest'}=\{ & \langle a_1,\dots,a_j\rangle: a_1,\dots,a_j\in A, \\
& \exists T\in\mathcal H, \{a_1,\dots,a_j\}=\dom(T), U(T,a_1,\dots,a_j)\subseteq U_n\}
\end{array}$$

We shall show that $\aest'$ satisfies the desired conditions:
\begin{itemize}
\item[(a)] By definition every relation $T$ in $\mathcal H$ is non-empty. Thus $(R_{\emptyset})^{\aest'}=\emptyset$.
\item[(b)] Let $a_1,\dots,a_k$ be (not necessarily different) elements in $A$ and let 
$1\leq l_1,\dots,l_j\leq k$. We shall show that $\langle a_{l_1},\dots,a_{l_j}\rangle\in(R_U)^{\aest'}$
where 
$$U=\{\langle h(a_{l_1}),\dots,h(a_{l_j})\rangle: h\in\hom(\aest_{\{a_1,\dots,a_k\}},\best)\}$$
Since $\mathcal H$ is an supercomplete  winning strategy, 
then the relation $$T=\{ h_{|\{a_{l_1},\dots,a_{l_j}\}}: h\in\hom(\aest_{|\{a_1,\dots,a_k\}},\best),h_{|\emptyset}\in\{\lambda\}\}$$
which is equivalent to 
$$\{h_{|\{a_{l_1},\dots,a_{l_j}\}}: h\in\hom(\aest_{|\{a_1,\dots,a_k\}},\best)\}$$
belongs to $\mathcal H$. In consequence we have $\langle a_{l_1},\dots,a_{l_j}\rangle\in(R_{U(T,a_{l_1},\dots,a_{l_j})})^{\aest'}$.
Finally, we have $U(T,a_{l_1},\dots,a_{l_j})=U$.

\item[(c)] Let $a_1,\dots,a_k$ be (not necessarily different) elements in $A$, let $R_{U_n}$ be a 
relation symbol, let $\langle a_{i_1},\dots,a_{i_j}\rangle\in(R_{U_n})^{\aest'}$ and let
$1\leq l_1,\dots,l_j\leq k$. We shall show that $\langle a_{l_1},\dots,a_{l_j}\rangle\in(R_U)^{\aest'}$
where 
$$U=\{\langle h(a_{l_1}),\dots,h(a_{l_{j}})\rangle:
h\in\hom(\aest_{\{a_1,\dots,a_k\}},\best),
\langle h(a_{i_1}),\dots,h(a_{i_j})\rangle\in U_n\}$$  
First, there exists some $T\in\mathcal H$ that has domain $\{a_{i_1},\dots,a_{i_j}\}$ and such that the inclusion
$U(T,a_{i_1},\dots,a_{i_j})\subseteq U_n$ holds. Since $\mathcal H$ is a 
supercomplete winning strategy we have that 
$$\begin{array}{ll}
T' & =\{h_{|\{a_{l_1},\dots,a_{l_j}\}}: h\in\hom(\aest_{|\{a_1,\dots,a_k\}},\best),h_{|\{a_{i_1},\dots,a_{i_j}\}}\in T\} \\
& \subseteq\{h_{|\{a_{l_1},\dots,a_{l_j}\}}: h\in\hom(\aest_{|\{a_1,\dots,a_k\}},\best),\langle h(a_{i_1}),\dots,h(a_{i_j})\rangle\in 
U_n\}\end{array}
$$ belongs also to $\mathcal H$. 

Thus,
$\langle a_{l_1},\dots,a_{l_j}\rangle$ belongs to $(R_{U(T',a_{l_1},\dots,a_{l_j})})^{\aest'}$ and it is
not difficult to see that $U(T',a_{l_1},\dots,a_{l_j})\subseteq U$.

\item[(d)] Straightforward from the definition.

\end{itemize}

For the converse, let $\aest'$ be a $\{R_{U_n}:1\leq n\leq m\}$-structure satisfying (a), (b), (c), and (d).
Let $\mathcal H$ be the collection of relations
$$\{\{\lambda\}\}\cup\{T:\exists a_1,\dots,a_j\in A,\dom(T)=\{a_1,\dots,a_j\},\langle a_1,\dots,a_j\rangle\in (R_{U(T,a_1,\dots,a_j)})^{\aest'}\}$$

We shall show that $\mathcal H$ is an supercomplete winning strategy for the Duplicator for the $(j,k)$-pebble-relation
game on $\aest$ and $\best$. It is immediate to show that $\mathcal H$ satisfies conditions (a), (b) and (d) of the
definition of a superstrict winning strategy. For condition (c), 
let $I\subseteq A$ with $|I|\leq k$, let $T$ be a relation in $\mathcal H$ with domain 
$I'\subseteq I$ and let $I''\subseteq I$ with $|I''|\leq j$. We have to show that
$T'=\{h_{|I''}:h\in\hom(\aest
_{|I},\best),h_{|I'}\in T\}$ belongs to $\mathcal H$.
We will do a case analysis. 

First assume that $I''=\emptyset$. In this case, since property (b) of the definition of supercomplete
winning strategy is satisfied we have that $T'=\{\lambda\}$ which belongs to $\mathcal H$.
Now assume that $I''\neq\emptyset$, say  $I=\{a_1,\dots,a_k\}$
(here $a_1,\dots,a_k$ are not necessarily different) 
and $I''=\{a_{l_1},\dots,a_{l_j}\}$
\begin{itemize}
\item First consider the case $I'=\emptyset$. Thus $T=\{\lambda\}$ and
$T'=\{h_{|I''}:h\in\hom(\aest_{|I},\best)\}$. From condition (b) we have that 
$\langle a_{l_1},\dots,a_{l_j}\rangle\in (R_U)^{\aest'}$ where 
$$U=\{\langle h(a_{l_1}),\dots,h(a_{l_{j}})\rangle:
h\in\hom(\aest_{\{a_1,\dots,a_k\}},\best)\}=U(T',a_{l_1},\dots,a_{l_j}),$$
and thus $T'\in\mathcal H$. 
\item The case $I'\neq\emptyset$ is proven similarly. In this case we have that there exist
some $a_{i_1},\dots,a_{i_j}\in I$ such that $I'=\{a_{i_1},\dots,a_{i_j}\}$ and  
$\langle a_{i_1},\dots,a_{i_j}\rangle\in (R_{U(T,a_{i_1},\dots,a_{i_j})})^{\aest'}$. 
Thus, $T'=\{h_{|I''}:h\in\hom(\aest_{|I},\best),h_{|I'}\in T\}$ is identical to
$$\{h_{|\{a_{l_1},\dots,a_{l_j}\}}:h\in\hom(\aest_{|\{a_1,\dots,a_k\}},\best),\langle h(a_{i_1}),\dots,h(a_{i_j})\rangle\in (R_{U(T,a_{i_1},\dots,a_{i_j})})^{\aest'}\}$$
From condition (c) of relational winning strategy we have that $\langle a_{i_1},\dots,a_{i_j}\rangle\in(R_U)^{\aest'}$
where 
$$\begin{array}{lll}
U= & \{\langle h(a_{l_1}),\dots,h(a_{l_j})\rangle: & h\in\hom(\aest_{|\{a_1,\dots,a_k\}},\best),\\
& & \langle h(a_{i_1}),\dots,h(a_{i_j})\rangle \in (R_{U(T,a_{i_1},\dots,a_{i_j}))})^{\aest'}\}
\end{array}$$
which is equal to $U(T',a_{l_1},\dots,a_{l_j})$; henceforth, $T'\in\mathcal H$.
\end{itemize}

\endproof

After the equivalence of supercomplete winning strategies and relational winning strategies has
been established, we shall construct a sentence $\varphi$ over the vocabulary $\vocab\cup\{=\}$
that tests whether such a structure $\aest'$ (certifying the existence of a relational winning strategy) exists. 
This is our third (and final) component of the proof of Theorem~\ref{the:expressivitygame}.

\begin{claim}
\label{cl:claim3}
For every $\best$ there exists a sentence $\varphi$ over the vocabulary $\vocab\cup\{=\}$ such
that for every $\vocab$-structure $\aest$, $\aest\models\varphi$ iff the Duplicator has a relational
winning strategy for the $(j,k)$-PR game on $\aest$ and $\best$.
\end{claim}

\proof
{\sloppypar The sentence $\varphi$ has a second-order predicate $R_{U_i}$ of arity
$j$ for every $j$-ary relation $U_i, 1\leq i\leq m$ over $B$. Thus
$$\varphi=\exists R_{U_1},\dots,R_{U_m}\forall v_1,\dots,v_k\psi(v_1,\dots,v_k),$$
%where $\psi(v_1,\dots,v_k)$ is a quantifier-free first-order formula
%over the vocabulary $\vocab\cup\{=,R_{U_1},\dots,R_{U_m}\}$, with
%variables among $v_1,\dots,v_k$.  Let us describe $\psi$. The formula
%$\psi$ will be a first-order formula in conjunctive normal form.
where $\psi(v_1,\dots,v_k)$ is a first-order formula
over the vocabulary $\vocab\cup\{=,R_{U_1},\dots,R_{U_m}\}$ that is
quantifier-free and has
variables among $v_1,\dots,v_k$.  Let us describe $\psi$. The formula
$\psi$ will be a first-order formula in conjunctive normal form.
First we need some auxiliary definitions, let $\vocab'$ be $\vocab\cup\{=,R_{U_1},\dots,R_{U_m}\}$
and let $\best'$ be the $\vocab'$-structure with universe $B$ such that for every $R\in\vocab$, 
$R^{\best'}=R^{\best}$, $(=)^{\best'}=\{(b,b):b\in B\}$, and such that for every $1\leq i\leq m$, $(R_{U_i})^{\best'}=U_i$.}

A disjunctive formula $\gamma(v_1,\dots,v_k)$ is any first-order quantifier-free formula obtained
as the disjunction of some (possibly negated) predicates in $\vocab'$ applied to variables in $v_1,\dots,v_k$.
$\gamma$ is monotone if every occurrence of a predicate in $\vocab'$ is negated.

Let $\gamma(v_1,\dots,v_k)$ be any disjunctive monotone formula over the vocabulary
$\vocab'$ with variables among $v_1,\dots,v_k$ and let $1\leq i_1,\dots,i_j\leq k$, 
be a collection of indices. We define $U(\gamma,i_1,\dots,{i_{j}})$ 
as the $j$-ary relation over $B$ defined by 
$$\{\langle b_{i_1},\dots,b_{i_j}\rangle: \best',b_1,\dots,b_k\not\models\gamma(v_1,\dots,v_k)\}$$ 
We are now in a position to describe $\psi(v_1,\dots,v_k)$. The formula $\psi$ is of the form 
$\bigwedge \Psi$ with $\Psi=\Psi_1\cup\Psi_2\cup\Psi_3$ where: 
\begin{itemize}
\item
$\Psi_1$ contains the formula
$\neg R_{\emptyset}(v_{l_1},\dots,v_{l_j})$ for every $1\leq l_1,\dots,l_j\leq k$.
\item
$\Psi_2$ contains for every disjunctive monotone formula $\gamma$ over
the vocabulary $\vocab'$ with at most one occurrence of a second-order predicate and
every collection of indices $i\leq i_i,\dots,i_j\leq k$, the formula (clause)
$\gamma(v_1,\dots,v_k)\vee R_{U(\gamma,i_1,\dots,i_{j})}(v_{i_1},\dots,v_{i_{j}})$. 
Notice that such clause is monotone and restricted Krom, although it might contain equalities
\item 
$\Psi_3$ contains for every $1\leq n,n'\leq m$ such that $U_n\subseteq U_n'$, and
for every $1\leq l_1,\dots,l_j\leq k$ the formula 
$\neg R_{U_n}(v_{l_1},\dots,v_{l_j})\vee R_{U_{n'}}(v_{l_1},\dots,v_{l_j})$.
\end{itemize}

Informally, each one of the subsets $\Psi_1,\Psi_2,\Psi_3$ of $\Psi$ encodes a condition of the definition
of relational winning strategy. Indeed, it is not difficult to see that $\bigwedge \Psi_1$ is equivalent
to condition (a) of relational winning strategy and that $\bigwedge \Psi_3$ formulates condition (d) of
relational winning strategy. It is also possible to see, although this case is certainly more
complicated, that $\bigwedge \Psi_2$ encodes exactly conditions (b) and (c) of relational winning strategy.
The intuition here is that the collection of all 
formulas $\gamma(v_1,\dots,v_k)\vee R_{U(\gamma,i_1,\dots,i_{j})}(v_{i_1},\dots,v_{i_{j}})$
in $\Psi$ where $\gamma$ does not contain a second-order predicate encodes (b) whereas
the set of all such formulas with $\gamma$ containing one occurrence of a second-order predicate encodes (c).

In the following we shall make all this more precise.

Let $\aest'$ be a $\vocab'$-structure. We shall show that $\aest'\models\forall v_1,\dots,v_k\psi(v_1,\dots,v_k)$
iff $\aest''=\aest'[\{R_{U_n}:1\leq n\leq m\}]$ is a relational winning strategy for the Duplicator
for the $(j,k)$-PR game on $\aest=\aest'[\vocab]$ and $\best$. It is easy to observe that this implies
our claim.

First, assume that $\aest'\models\forall v_1,\dots,v_k\psi(v_1,\dots,v_k)$. Thus, for every $a_1,\dots,a_k$
in $A$, $\aest',a_1,\dots,a_k\models\psi(v_1,\dots,v_k)=\bigwedge\Psi$. Since for every 
$1\leq l_1,\dots,l_j\leq k$, $\neg R_{\emptyset}(v_{l_1},\dots,v_{l_j})$ is in $\Psi$ we have that
$\langle a_{l_1},\dots,a_{l_j}\rangle\not\in (R_{\emptyset})^{\aest''}$. Thus 
$(R_{\emptyset})^{\aest''}=\emptyset$ and,
in consequence, $\aest''$ satisfies condition (a) on the definition of relational winning strategy.

We shall show that $\aest''$ satisfies (b). Let $a_1,\dots,a_k$ be (not necessarily different) elements
of $A$. 
Let $\theta(v_1,\dots,v_k)=\canonic(\aest,a_1,\dots,a_k)(v_1,\dots,v_k)$.

Thus, we have $\aest',a_1,\dots,a_k\models\neg\theta(v_1,\dots,v_k)\vee R_{U(\neg\theta,l_1,\dots,l_j)}(v_{l_1},\dots,v_{l_j})$. 
Since by Proposition~\ref{pro:canonic},  $\aest',a_1,\dots,a_k\not\models\neg\theta(v_1,\dots,v_k)$ we have that
$\langle a_{l_1},\dots,a_{l_j}\rangle\in (R_{U(\neg\theta,l_1,\dots,l_j)})^{\aest''}$.
Finally we have:
$$\begin{array}{ll}
 U(\neg\theta,l_1,\dots,l_j) & =\{\langle b_{l_1},\dots,b_{l_j}\rangle:\best',b_1,\dots,b_k\not\models\neg\theta(v_1,\dots,v_k)\} \\
& =\{\langle b_{l_1},\dots,b_{l_j}\rangle: \aest_{|\{a_1,\dots,a_k\}},a_1,\dots,a_k\maps\best',b_1,\dots,b_k\} \\
& =\{\langle h(a_{l_1}),\dots,h(a_{l_j})\rangle: h\in\hom(\aest_{|\{a_1,\dots,a_k\}},\best)\}
\end{array}$$

We shall show that $\aest''$ satisfies condition (c). Let $a_1,\dots,a_k$ be (not necessarily different) elements
of $A$, and let $\theta$ be defined
as above. Let $R_{U_n}$ be a relation symbol, let $\langle a_{i_1},\dots,a_{i_j}\rangle\in (R_{U_n})^{\aest''}$
and let $1\leq l_1,\dots,l_j\leq k$.
Let $\gamma(v_1,\dots,v_k)$ be the disjunctive monotone formula defined by 
$\gamma(v_1,\dots,v_k)=\neg\theta(v_1,\dots,v_k)\vee\neg R_{U_n}(v_{i_1},\dots,v_{i_j})$. 
Therefore the formula $\gamma(v_1,\dots,v_k)\vee (R_{\gamma,l_1,\dots,l_j})(v_{l_1},\dots,v_{l_j})$ is in $\Psi$. Thus,
$\aest'',a_1,\dots,a_k\models (R_{U(\gamma,l_1,\dots,l_j)})(v_{l_1},\dots,v_{l_j})$ and in consequence 
$\langle a_{l_1}\dots,a_{l_j}\rangle\in (R_{U(\gamma,l_1,\dots,l_j)})^{\aest''}$.
Finally we have:
$$\begin{array}{lll}
U(\gamma,l_1,\dots,l_j) & =\{ & \langle b_{l_1}\dots,b_{l_j}\rangle:\best',b_1,\dots,b_k\not\models\gamma(v_1,\dots,v_k)\} \\
& =\{ & \langle b_{l_1}\dots,b_{l_j}\rangle:\best',b_1,\dots,b_k\not\models\neg\theta(v_1,\dots,v_k), \\
& & \langle b_{i_1},\dots,b_{i_j}\rangle\in (R_{U_n})^{\best'}=U_n\} \\
& =\{ & \langle h(a_{l_1}),\dots,h(a_{l_j})\rangle:h\in\hom(\aest_{|\{a_1,\dots,a_k\}},\best), \\
& & \langle h(a_{i_1}),\dots,h(a_{i_j})\rangle\in U_n\}
\end{array}$$
We shall show that $\aest''$ satisfies condition (d). Let $1\leq n,n'\leq m$ such that $U_n\subseteq U_{n'}$,
and let $\langle a_{l_1},\dots,a_{l_j}\rangle\in (R_{U_n})^{\aest''}$. Since 
$\neg R_{U_n}(v_{l_1},\dots,v_{l_j})\vee R_{U_{n'}}(v_{l_1},\dots,v_{l_j})$ is in $\Psi_3$ we have that
$\aest',a_1,\dots,a_k\models \neg R_{U_n}(v_{l_1},\dots,v_{l_j})\vee R_{U_{n'}}(v_{l_1},\dots,v_{l_j})$
and, in consequence, $\langle a_{l_1},\dots,a_{l_j}\rangle\in(R_{U_{n'}})^{\aest''}$.

For the converse, assume that $\aest''$ is a relational winning strategy. We shall show that 
for all $a_1,\dots,a_k$ (not necessarily different) elements in $A$ and for every
$\chi(v_1,\dots,v_k)$ in $\Psi$, $\aest',a_1,\dots,a_k\models\chi(v_1,\dots,v_k)$. First,
if $\chi\in\Psi_1$ then $\chi=\neg R_{\emptyset}(v_{l_1},\dots,v_{l_j})$ for some $1\leq l_1,\dots,l_j\leq k$.
In this case $\aest',a_1,\dots,a_k\models\chi(v_1,\dots,v_k)$ since 
$\langle a_{l_1},\dots,a_{l_j}\rangle\not\in (R_{\emptyset})^{\aest''}=\emptyset$. Secondly, if $\chi\in\Psi_2$
then we have $\chi(v_1,\dots,v_k)=\gamma(v_1,\dots,v_k)\vee R_{U(\gamma,l_1,\dots,l_j)}(v_{l_1},\dots,v_{l_j})$ 
where $\gamma(v_1,\dots,v_k)$ is a disjunctive monotone formula with at most one occurrence of
a predicate in $\{R_{U_n}:1\leq n\leq m\}$. Assume first that $\gamma(v_1,\dots,v_k)$ is a formula over $\vocab\cup\{=\}$
(that is, it does not contain a second-order predicate).
Thus, if $\aest,a_1,\dots,a_k\not\models\gamma(v_1,\dots,v_k)$ implies the following. Let 
$\theta(v_1,\dots,v_k)=\canonic(\aest,a_1,\dots,a_k)(v_1,\dots,v_k)$. Then, by Proposition~\ref{pro:canonic}, 
$\theta(v_1,\dots,v_k)$ implies $\neg\gamma(v_1,\dots,v_m)$.

Thus 
$$\begin{array}{ll}
U(\gamma,l_1,\dots,l_j) & =\{\langle b_{l_1},\dots,b_{l_j}\rangle: \best',b_1,\dots,b_k\not\models\gamma(v_1,\dots,v_k)\} \\
& \supseteq\{\langle b_{l_1},\dots,b_{l_j}\rangle: \best',b_1,\dots,b_k\models\theta(v_1,\dots,v_k)\} \\
& =\{\langle b_{l_1},\dots,b_{l_j}\rangle :\aest_{\{a_1,\dots,a_k\}},a_1,\dots,a_k\maps\best,b_1,\dots,b_k \} \\
& =\{\langle h(a_{l_1}),\dots,h(a_{l_j}): h\in\hom(\aest_{|\{a_1,\dots,a_k\}},\best)\}
\end{array}
$$

Let $U$ be $\{\langle h(a_{l_1}),\dots,h(a_{l_j}): h\in\hom(\aest_{|\{a_1,\dots,a_k\}},\best)\}$.
Since $\aest''$ is a relational winning strategy, then 
$\langle a_{l_1}\dots,a_{l_j}\rangle\in (R_U)^{\aest''}$. Furthermore, since $U\subseteq U(\gamma,l_1,\dots,l_j)$
we have that $\langle a_{l_1},\dots,a_{l_j}\rangle\in (R_{U(\gamma,l_1,\dots,l_j)})^{\aest''}$.

The proof when $\gamma$ contains a predicate in $\{R_{U_n}:1\leq n\leq m\}$ is analogous.
Finally it is straightforward to verify the case $\chi\in\Psi_3$ using condition (d) of relational winning strategy.

\endproof

Finally, the proof of Theorem~\ref{the:expressivitygame} is a consequence of Claims~\ref{cl:claim1}, ~\ref{cl:claim2},
and~\ref{cl:claim3}. 
\endproof

Restricted Krom SNP formulas can be regarded alternatively as a particular type of Datalog programs
called, linear Datalog Programs.

Let $\vocab$ be a vocabulary.
A {\em Datalog Program} over $\vocab$ is a finite set of rules of the form
$$t_0  \; \text{:--} \;\; t_1,\dots,t_m$$
where each $t_i$ is an atomic formula $R(v_1,\dots,v_m)$. The relational predicates that occur in the 
heads of the rules are the {\em intensional database} predicates (IDBs) and do not belong to $\vocab$,
while all others are the
{\em extensional database} predicates (EDBs) and must belong to $\vocab$. One of the IDBs is designated 
as the {\em goal} of the program. Note that IDBs may occur in the bodies of rules, and, thus, a Datalog
program is a recursive specification of the IDBs with semantics obtained via least fixed-points of
monotone operators (see~\cite{Ullman:1989}).

A Datalog Program is called linear if every rule contains at most one occurrence of a IDB in its body.
Let $0\leq j\leq k$ be non-negative integers, $(j,k)$-Datalog is said to be the collection of all Datalog programs
in which every rule has at most $k$ variables and at most $j$ variables in the head.

Let us introduce with a bit more detail the semantics of Datalog Programs.

%Let $Q$ be a Datalog Program, let $\vocab_{\text{EDB}}$ be the set of its extensional predicates,
%let $\vocab_{\text{IDB}}$ be the set of its intensional predicates, let $\vocab$ be any vocabulary
%containing $\vocab_{\text{EDB}}\cup\vocab_{\text{IDB}}$ (it is usual to consider only the
%vocabulary $\vocab=\vocab_{\text{EDB}}\cup\vocab_{\text{IDB}}$ but we have some built-in in mind).

Let $Q$ be a Datalog Program over $\vocab$, let $\vocab_{\text{IDB}}$ be the set of its intensional predicates, and let
$\vocab'=\vocab\cup\vocab_{\text{IDB}}$. 
The Datalog Program $Q$ defines an operator $\Phi:\str[\vocab']\rightarrow\str[\vocab']$ 
in the class of $\vocab'$-structures in the following way: For every $\vocab'$-structure $\aest$, $\Phi(\aest)$ is defined to be the smallest
$\vocab'$-structure with the same universe $A$ of $\aest$ such that: 
\begin{itemize}
\item for every $R\in\vocab'$, $R^{\aest}\subseteq R^{\Phi(\aest)}$,
\item for every rule $t_0 \; \text{:-} \;\; t_1,\dots,t_m$ in $Q$, with variables $u_1,\dots,u_n$, and for every interpretation $h:\{u_1,\dots,u_n\}\rightarrow A$ 
such that $$\aest,h(u_1),\dots,h(u_n)\models t_1\wedge \cdots \wedge t_m,$$
we have that $\langle h(v_1),\dots,h(v_r)\rangle\in P^{\Phi(\aest)}$ where $t_0=P(v_1,\dots,v_r)$
\end{itemize}

Since $\Phi$ is a monotone operator we can define its minimum fix point $\Phi^*:\str[\vocab']\rightarrow\str[\vocab']$ as:
$$\Phi^*(\aest)=\bigcup_{n\geq 0} \Phi^n(\aest), \text{ for every } \aest$$
where $\Phi^0(\aest)=\aest$ and $\Phi^{n+1}(\aest)=\Phi(\Phi^n(\aest)), n\geq 0$. 
Alternatively, $\Phi^*(\aest)$ can be defined as the smallest $\vocab'$-structure with universe $A$ such
that $\aest\subseteq\Phi^*(\aest)$ and $\Phi(\Phi^*(\aest))=\Phi^*(\aest)$.

A distinguished IDB predicate $P$ is designed to be the goal of the program. A Datalog
program is a query that given a $\vocab$-structure $\aest$, 
returns $P^{\Phi^*(\aest')}$, where
$\aest'$ is the $\vocab'$-structure with the same universe of $\aest$ and
such that for all $R\in\vocab$, $R^{\aest'}=R^{\aest}$
and for all $R\in\vocab_{IDB}$, $R^{\aest'}=\emptyset$.
We say that
a structure $\aest$ is accepted by the Datalog Program $Q$ iff $P^{\Phi^*(\aest')}\neq\emptyset$.

Let $\mathcal C$ be a set of $\vocab$-structures and let $Q$ be a Datalog Program. We
say that $\mathcal C$ is defined by $Q$ if for every $\vocab$-structure $\aest$, 
$\aest$ is in $\mathcal C$ iff $\aest$ is accepted by $Q$. It is well-known and
easy to see that if $\mathcal C$ is accepted by a Datalog Program then $\mathcal C$ must be a filter.

The following result relates linear Datalog programs with infinitary logics. 

\begin{theorem}
\label{the:dataloginpathlogic}
Let $0\leq j\leq k$ be non-negative integers, and let $\mathcal C$ be a set of $\vocab$-structures
for some vocabulary $\vocab$. If $\mathcal C$ is definable by a linear $(j,k)$-Datalog Program
then it is also definable in $\pathlogic^{j,k}$.
\end{theorem}
\proof

Let $Q$ be a linear $(j,k)$-Datalog Program. We will show that: (*) for every $\vocab$-structure 
$\aest$, for every intensional predicate $R\in\vocab_{\text{IDB}}$, 
for every $n\geq 0$ and for every $\langle a_1,\dots,a_r\rangle\in R^{\Phi^n(\aest')}$, 
there exists some structure $\best$, some $b_1,\dots,b_r\in B$, some path-decomposition $S_1,\dots,S_s$ of width $(j,k)$
of $\best$ with $\{b_1,\dots,b_r\}\subseteq S_1$, such that $\langle b_1,\dots,b_r\rangle\in R^{\Phi^n(\best')}$, and
$\best,b_1,\dots,b_r\maps\aest,a_1,\dots,a_r$.  We will prove statement
(*) by induction on $n$. The statement is vacuously true for $n=0$. We will show that (*) holds
for $n+1$ whenever it holds for $n$. Let $R\in\vocab_{\text{IDB}}$ be a intensional predicate
and let $\langle a_1,\dots,a_r\rangle$ be any tuple in $R^{\Phi^{n+1}(\aest')}$. If $\langle a_1,\dots,a_r\rangle$ is
in $R^{\Phi^n(\aest')}$ then we are done. Otherwise there exists some rule
$R(y_1,\dots,y_r)\; \text{:-} \; \; t_1,\dots,t_m$ in $Q$, over the variables $u_1,\dots,u_{k'}$, $k'\leq k$ and
some mapping $h:\{u_1,\dots,u_{k'}\}\rightarrow A$ such
that $\Phi^n(\aest'),h(u_1),\dots,h(u_{k'})\models t_1\wedge\cdots\wedge t_m$,
and $a_i=h(y_i)$, for every $1\leq i\leq r$. Let us consider two cases: if the
body of the rule does not contain any intensional predicate, then
$\aest,h(u_1),\dots,h(u_{k'})\models t_1\wedge\cdots\wedge t_m$. Then,
$\aest_{|\{h(u_1),\dots,h(u_{k'})\}}$ and $h(a_1),\dots,h(a_{k'})$ satisfy statement 
(*). Otherwise, let us assume that the body contains one occurrence of an intensional
predicate, say $t_1=R_1(x_1,\dots,x_{l})$.  In this case, since 
$$\langle h(x_1),\dots,h(x_l)\rangle\in R_1^{\Phi^n(\aest')},$$
we can assume, by induction hypothesis, that there exists some 
$\best$ and some $b_1,\dots,b_l$ in $B$, such
that $\langle b_1,\dots,b_r\rangle\in R^{\Phi^*(\best')}$
and 
$\best,b_1,\dots,b_l\maps\aest,h(x_1),\dots,h(x_l)$.
Furthermore there exists a path-decomposition $S_1,\dots,S_s$ of $\best$
such that $\{b_1,\dots,b_l\}\subseteq S_1$. We can assume,
by renaming elements of $B$ if necessary, 
that $b_i=h(x_i)$ for every $1\leq i\leq l$ and
that every other element of the universe of $B$ does
not belong to $\aest$. Consider the structure 
$\cest$ given as $\best\cup\aest_{|\{h(u_1),\dots,h(u_{k'})\}}$.
By construction, $\Phi^n(\cest'),h(u_1)\dots,h(u_{k'})\models t_1\wedge\cdots\wedge t_m$
and consequently, $\langle h(x_1),\dots,h(x_l)\rangle\in R^{\Phi^{n+1}(\cest')}$.

Furthermore $\{h(u_1),\dots,h(u_{k'})\},S_1,\dots,S_s$ is a path-decomposition
of $\cest$ of width $(j,k)$.

In order to finish the proof, let $\mathcal O$ be the set containing all
$\vocab$-structures of pathwidth $(j,k)$ in $\mathcal C$. We shall
see that $\mathcal O$ is an obstruction set of $\neg \mathcal C$.
First, as $\mathcal O$ only contains structures in $\mathcal C$ 
and $\mathcal C$ is a filter, we
can infer that for every structure $\aest$, such that $\best\maps\aest$
for some $\best$ in $\mathcal O$, $\aest$ is in $\mathcal C$.
Conversely, let $\aest$ be any structure in $\mathcal C$. By the definition of
acceptance by a Datalog Program, we can conclude that $P^{\Phi^*(\aest')}\neq\emptyset$.
By (*) we have that there exists some $\vocab$-structure $\best$ with pathwidth at most $(j,k)$ 
such that $P^{\Phi^*(\best')}\neq\emptyset$ (and hence $\best\in \mathcal O$), 
and $\best\maps\aest$. 

\endproof

\begin{lemma}
\label{le:equivdatalogSNP}
Let $0\leq j\leq k$ be non-negative integers and let $\mathcal C$ be a collection of $\vocab$-structures.
The two following sentences are equivalent:
\begin{enumerate}
\item $\mathcal C$ is definable in linear $(j,k)$-Datalog.
\item $\neg\mathcal C$ is definable in $j$-adic $k$-ary restricted Krom monotone SNP.
\end{enumerate}
Furthermore, if $\neg\mathcal C$ is an ideal we also have that (1) and (2) are equivalent to:
\begin{itemize}
\item[(3)] $\neg\mathcal C$ is definable in $j$-adic $k$-ary restricted Krom SNP with equalities.
\end{itemize}
\end{lemma}
\proof
To show the equivalence between {\bf (1)} and {\bf (2)} is straightforward but laborious.
Here we will only sketch briefly the proof. To proof \imply{1}{2} we show that
for any any linear $(j,k)$-Datalog program $Q$, there exists a sentence $\varphi$
in $j$-adic $k$-ary restricted Krom monotone SNP, such that for every structure $\aest$,
$\aest\models\varphi$ iff and only if $Q$ does not accept $\aest$ . 
The sentence $\varphi$ is of
the form $\exists R_1,\dots,R_l \forall v_1,\dots,v_k \psi(v_1,\dots,v_k)$ where:
\begin{itemize}
\item $R_1,\dots,R_l$ are the IDBs of the Datalog Program $Q$
\item $v_1,\dots,v_k$ are the variables occurring in the rules of the Datalog Program. Note: We
can assume that the variables in the Datalog Program have been renamed so that every rule
has its variables among $v_1,\dots,v_k$
\item $\psi(v_1,\dots,v_k)$ is a CNF formula that is defined as
$$\psi(v_1,\dots,v_k) = \neg P\wedge\bigwedge_{t_0 \; \text{:-} \;\; t_1,\dots,t_m\in Q} t_0\vee\neg t_1\vee\cdots\vee\neg t_m$$
where $P$ is the goal predicate of $Q$.
\end{itemize}
It is immediate, although again laborious, to show that $\varphi$ is a sentence in $j$-ary
$k$-ary restricted Krom monotone SNP that is satisfied precisely by those
structure that are not accepted by $Q$.

For the converse (\imply{2}{1}), let 
$\varphi=\exists R_1,\dots,R_l\forall v_1,\dots,v_k \psi(v_1,\dots,v_k)$ be an arbitrary sentence
in $j$-adic $k$-ary restricted Krom monotone SNP. We shall construct a linear $(j,k)$-Datalog
Program $Q$ in the following way:
\begin{itemize}
\item The IDBs of $Q$ are precisely the second order predicates of $\varphi$, $R_1,\dots,R_l$ 
plus a new $0$-ary IDB, $P$, which will act as the goal predicate.
\item The rules of $Q$ are constructed from $\psi$ in the following way. The formula $\psi$ is a CNF 
and henceforth is the conjunction of several subformulas, where each one of this subformulas
is the disjunction of atomic or negated atomic formulas. In fact, since we
are dealing with a restricted Krom monotone SNP formula each one of this subformulas has
to be of a very restricted form. More precisely, every one of the subformulas that 
constitute $\psi$ has
to be either the disjunction of negated atomic formulas $\neg t_1\vee\cdots\vee\neg t_m$,
or it can contain one unnegated atomic formula, that is, it has to be of the form 
$t_o\vee\neg t_1\vee\cdots\vee\neg t_m$ where the underlying predicate of $t_0$ is
an IDB. 

Datalog Program $Q$ contains a rule per each subformula in $\psi$. If
the subformula is of the form $\neg t_1\vee\cdots\vee\neg t_m$ then the rule
added to $Q$ is $P\; \text{:-} \;\; t_1,\dots,t_m$. Otherwise, if the subformula is of the
form $t_0\vee\neg t_1\vee\cdots\vee\neg t_m$ then the rule is of the form 
$t_0 \; \text{:-} \; \; t_1,\dots,t_m$. 
\end{itemize}
Again it is an easy exercise to show that the Datalog Program $Q$ defined is
indeed a linear $(j,k)$-Datalog Program that accepts precisely those structures
that falsify $\varphi$.

To see that if $\mathcal C$ is a filter then {\bf (3)} implies {\bf (1)} we make use of a result by Feder and Vardi~\cite{Feder/Vardi:03}
which states that for every Datalog($\neq$,$\neg$) Program $P$, that is, every Datalog Program in which
we allow inequality and the negation of EDBs, that defines a filter $\mathcal C$, there exists a Datalog Program $P'$,
that defines the same set of structures $\mathcal C$. Furthermore, a closer inspection to the proof in~\cite{Feder/Vardi:03} 
reveals that if $P$ is in linear $(j,k)$-Datalog($\neq$,$\neg$), then $P'$ is
in linear $(j,k)$-Datalog..

The proof that ${\bf (3)}$ implies ${\bf (1)}$ mimics that of ${\bf (2)}$ implies ${\bf (1)}$. As before
let  $\varphi=\exists R_1,\dots,R_l\forall v_1,\dots,v_k \psi(v_1,\dots,v_k)$ be an arbitrary sentence
in $j$-adic $k$-ary restricted Krom SNP with equalities. We shall construct a linear $(j,k)$-Datalog
Program $Q$ in a similar fashion:
\begin{itemize}
\item The IDBs of $Q$ are precisely the second order predicates of $\varphi$, $R_1,\dots,R_l$ 
plus a new $0$-ary IDB, $P$, which will act as the goal predicate.
\item The rules of $Q$ are constructed from $\psi$ in the following way. The formula $\psi$ is a CNF 
and henceforth is the conjunction of several subformulas of the form $t_1\vee\cdots\vee t_m$ , where each $t_i$, $1\leq i\leq m$ is 
an atomic or a negated atomic formula. Since the formula $\psi$ is not
supposed to be monotone and might contain equalities we only can assume that each one of the disjunctions 
$t_1\vee\cdots\vee t_m$ of $\psi$ has at most one positive
occurrence of a second-order variable and at most one negative occurrence of a second order-variable. 
As before, we add to $Q$ a rule per each disjunction $t_1\vee\cdots\vee t_m$ in $\psi$. If $t_1\vee\cdots\vee t_m$
does not contain any positive occurrence of an IDB, then the rule added to $Q$ is $P\; \text{:--} \;\; \neg t_1,\dots,\neg t_m$.
Otherwise, if $t_1\vee\cdots\vee t_m$ contains one positive occurrence of an IDB, say $t_1$, then the
rule added to $Q$ is $t_1\; \text{:--} \;\; \neg t_2,\dots,\neg t_m$. 
\end{itemize}
Observe that the body of a rule in $Q$
might contain equalities and negated atomic formulas $\neg R(x_1,\dots,x_r)$ where $R$ is an EDB.
Consequently, $Q$ is a Datalog($\neq$,$\neg$) Program. It is easy to see that
$Q$ is, indeed, a linear $(j,k)$-Datalog($\neq$,$\neg$) Program and that $Q$ accepts accepts precisely those structures
that falsify $\varphi$. Since $\mathcal C$ is a filter, by the result in~\cite{Feder/Vardi:03} mentioned above there
exists a linear $(j,k)$-Datalog Program (that is, without inequalities and negations) that defines $\mathcal C$. 

The implication (\imply{2}{3}) is trivial.
\endproof

If we are dealing with set of structures $\neg\mathcal C$ of the form $I(\best)$ for some $\vocab$-structure $\best$, then we can combine
Theorem~\ref{the:gamesexpressivity}, Lemma~\ref{le:equivdatalogSNP},
Theorem~\ref{the:expressivitygame}, and Theorem~\ref{the:dataloginpathlogic}, obtaining the main result of this paper.

\begin{theorem}
\label{the:main}
Let $0\leq j\leq k$ be non-negative integer,
let $\best$ be a $\vocab$-structure, and let $\neg\mathcal C=\csp(\best)=I(\best)$.
The following sentences are equivalent:
\begin{enumerate}
\item The class $\mathcal C$ is $\pathlogic^{j,k}$-definable.
\item The class $\mathcal C$ is $\restpathlogic^{j,k}$-definable.
\item The class $\mathcal C$ is definable in linear $(j,k)$-Datalog.
\item If $\aest$ and $\best$ are finite structures such that $\aest\in\mathcal C$ and
Duplicator has a winning strategy for the $(j,k)$-PR game on $\aest$ and $\best$, then 
$\best\in\mathcal C$.
\item The class $\neg\mathcal C$ has an obstruction set with pathwidth at most $(j,k)$.
\item The class $\neg\mathcal C$ is definable in $j$-adic $k$-ary restricted Krom SNP with equalities.
\item The class $\neg\mathcal C$ is definable in $j$-adic $k$-ary restricted Krom monotone SNP.
\end{enumerate}
\end{theorem}

Let $\best$ be a $\vocab$ structure. If $\neg\mathcal C=I(\best)$ satisfies any of the conditions
of Theorem~\ref{the:main}
we say that $\best$ has  {\em $(j,k)$-path duality}. Finally, we say
that $\best$ has {\em bounded path duality} if $\best$ has $(j,k)$-path duality for some $0\leq j<k$.

\section{Applications in Computational Complexity}

In this section we bring computational complexity into the picture. We shall start by proving
that for every finite structure $\best$ with bounded path duality, $\csp(\best)$ is in NL.

This result is an immediate consequence of the following theorem, proven in~\cite{Gradel:1992}.

\begin{theorem}[\cite{Gradel:1992}]
Let $\varphi$ be a Krom CNF formula over a vocabulary $\vocab$. The problem of
deciding, given a $\vocab$-structure $\aest$, whether $\aest\models\varphi$, is solvable in NL.
\end{theorem}

Indeed, if $\best$ has bounded path duality, then $\csp(\best)$ is definable in Krom SNP. Hence
we have,

\begin{proposition}
\label{pro:NL}
Let $\best$ be a $\vocab$-structure with bounded path duality. Then $\csp(\best)$ is in NL.
\end{proposition}

In what follows, we shall revisit all families of constraint satisfaction problems, that
up to the best of our knowledge, are known to be in NL. Our goal here is to show that
the notion of bounded path duality provides a unifying framework that encompasses and explains
this results. In the last part of this section, we shall see how
the notion of bounded path duality can be used to place new problems in NL.

\subsection{Implicational constraints}
\label{sec:implicational}

The class of {\em implicational constraints} was introduced independently by Kirousis~\cite{Kirousis:1993}
and by Cooper et al. ~\cite{Cooper/Cohen/Jeavons:1994} (in this second reference they were
named {\em 0/1/all constraints}). A binary relation $R\subseteq A^2$ over a finite domain $A$ is called implicational 
if it has one of the following three forms: (1) $B\times C$ for some $B,C\subseteq A$, (2) 
$\{(a,f(a)) :  a\in B\}$
for some $B\subseteq A$ and some one-to-one mapping $f:B\rightarrow A$, or (3) 
$(\{b\}\times C)\cup (B\times\{c\})$
for some $B,C\subseteq A$, $b\in B$, and $c\in C$. A structure $\best$ is said to be implicational
if so are all its relations. It is easy to observe that 2-SAT can be encoded as a constraint satisfaction
problem $\csp(\best_{\text{2-SAT}})$, with $\best_{\text{2-SAT}}$ implicational:

The signature $\vocab$ of $\best_{\text{2-SAT}}$ contains three binary relation symbols $P_0$, $P_1$ and $P_2$. The universe
of $\best$ is $\{0,1\}$ and the values of $P_0^{\best_{\text{2-SAT}}}$, $P_1^{\best_{\text{2-SAT}}}$, and $P_2^{\best_{\text{2-SAT}}}$ 
are
respectively $\{0,1\}^2/\{(0,0)\}$, $\{0,1\}^2/\{(0,1)\}$, and $\{0,1\}^2/\{(1,1)\}$. Observe that each
one of these relations is implicational.

It is well
known and quite easy to see that $2$-SAT and $\csp(\best_{\text{2-SAT}})$ are logspace reducible to each other.
Indeed, let $\varphi$ be any arbitrary $2$-CNF formula. We shall show that it is possible to construct 
in logarithmic space a $\vocab$-structure $\aest$ such that $\aest$ is homomorphic to $\best_{\text{2-SAT}}$ if and only if 
$\varphi$ has a solution. The universe $A$ of $\aest$ is given by the variables of $\varphi$. Furthermore, $P_0^{\aest}$
contains for each clause $(x\vee y)$ with only positive literals, the tuple $(x,y)$. Observe that every homomorphism
from $\aest$ to $\best$ should set values to $x$ and $y$ that belong to $P_0^{\best_{\text{2-SAT}}}=\{0,1\}^2/\{0,0\}$.
Hence the values set to $x$ and $y$ must satisfy clause $(x\vee y)$. 

A similar processing is applied to the other types
of clauses. More precisely, for each clause with
exactly one positive literal $(x\vee\neg y)$ (let us assume that the literals are ordered so that the positive
literal comes first) we add the tuple $(x,y)$ to $P_1^{\aest}$ and for each clause $(\neg x\vee\neg y)$ with only
negative literals we add the tuple $(x,y)$ to $P_2^{\aest}$. It is very easy to verify that satisfying assignments
of $\varphi$ correspond to homomorphisms from $\aest$ to $\best_{\text{2-SAT}}$ and that this reduction can
be performed in logarithmic space. The reduction from $\csp(\best_{\text{2-SAT}})$ to 2-SAT is also very simple and we
omit it.

Let $\best$ be any implicational structure and let $\aest$ be any instance of $\csp(\best)$. We first recall
a well-known result about constraint satisfaction problems $\csp(\best)$ with $\best$ implicational 
which can be found in~\cite{Kirousis:1993}. If $\aest$ is an instance of $\csp(\best)$ we define
the conflict graph $G$ associated to $\aest$ and $\best$ as the digraph $G=(V,E)$, with set of
nodes $V=\{(a,b) : a\in A,b\in B\}\cup\{\square\}$ and set of edges $E$ constructed in the following way:
\begin{itemize}
\item[(a)] For every predicate $R$ in the vocabulary $\vocab$ of $\aest$, for every
$b,b'\in B$ such that $R^{\best}\cap(\{b\}\times B)=\{(b,b')\}$, and for every tuple $(a,a')\in R^{\aest}$
we add to $E$ an arc from $(a,b)$ to $(a',b')$.
\item[(b)] For every predicate $R$ in the vocabulary $\vocab$ of $\aest$, for every
$b,b'\in B$ such that $R^{\best}\cap(B\times\{b'\})=\{(b,b')\}$, and for every tuple $(a,a')\in R^{\aest}$
we add to $E$ an arc from  $(a',b')$ to $(a,b))$.
\item[(c)] For every predicate $R$ in the vocabulary $\vocab$ of $\aest$, for every 
$b\in B$ such that $R^{\best}\cap(\{b\}\times B)=\emptyset$, and for every tuple $(a,a')\in R^{\aest}$
we add to $E$ an arc from $(a,b)$ to $\square$.
\item[(d)] For every predicate $R$ in the vocabulary $\vocab$ of $\aest$, for every 
$b'\in B$ such that $R^{\best}\cap(B\times\{b'\})=\emptyset$, and for every tuple $(a,a')\in R^{\aest}$
we add to $E$ an arc from $(a',b')$ to $\square$.
\end{itemize}

\begin{lemma}
\label{le:implicational}
$\aest$ is homomorphic to $\best$ iff for every $a\in A$ there exists some $b\in B$ such that there is not a path
from $(a,b)$ to any some node in $\{(a,b') : b'\in B\\ \{b\}\}\cup\{\square\}$.
\end{lemma}

Let us see that this implies that each implicational $\best$ has $(2,3)$-path duality. 

\begin{lemma}
\label{le:implicational2}
Every implicational structure $\best$ has $(2,3)$-path duality.
\end{lemma}
\proof
Let $\aest$ be any structure
not homomorphic to $\best$ and let $G$ be the conflict graph associated to $\aest$ and $\best$.
Let $\{b_1,\dots,b_m\}$ be the universe $B$ of $\best$. 
We shall
construct a structure $\pest$ with pathwidth at most $(2,3)$ that is homomorphic to $\aest$
but is not homomorphic to $\best$. 

By Lemma~\ref{le:implicational} there exists
some $a^*\in A$ such that for all $b_i$, $1\leq i\leq m$, there is a path in $G$
from $(a^*,b_i)$ to either $(a^*,b_j)$ for some $j\neq i$ or $\square$. 
Let us denote by $(a^*,b_i)=v^i_1,v^i_2,\dots,v^i_{l_i}$ the elements of such path. 
Observe that $v^i_{l_i}$ is either $(a^*,b_j)$ for some $j\neq i$ or $\square$.

The universe $P$ of $\pest$ contains for each $1\leq i\leq m$, and
for each node $v^i_k$, $1\leq k\leq l_i$ in the path corresponding to $b_i$
an element $w^i_k$. If the same node of $A$ appears in several paths
then we make different copies of it. Finally we merge nodes 
$w^i_k$, corresponding to $v^i_k=(a^*,b_l)$ for some $1\leq l\leq m$ into a unique
node that we shall call $w$. Observe
for all for all $1\leq i\leq m$,
$w^i_1$ is $w$. 

Now we shall
construct the relations of $\pest$ and, at the same time, a homomorphism
$h$ from $\pest$ to $\aest$. 

Let $i$ be any integer
with $1\leq i\leq m$ and let $v^i_j$, $v^i_{j+1}$ be any two consecutive
elements in the path corresponding to $b_i$. Then, $(v^i_j,v^i_{j+1})$ is an arc of $G$ that
must have been added according to $(a)$, $(b)$, $(c)$ or $(d)$. 
If $(v^i_j,v^i_{j+1})$ has
been added according to $(a)$ then there exists some predicate $R$
in $\vocab$, some $b,b'\in B$ such that $R^{\best}\cap(\{b\}\times B)=\{(b,b')\}$,
and some tuple $(a,a')\in R^{\aest}$. Then we add to $R^{\pest}$ the 
tuple $(w^i_{j+1},w^i_{j+1})$. The mapping $h$ maps $v^i_{j}$ to
$a$ and $v^i_{j+1}$ to $a'$. Similarly, if $(v^i_j,v^i_{j+1})$ has
been added according to $(b)$ then there exists some predicate $R$
in $\vocab$, some $b,b'\in B$ such that $R^{\best}\cap(B\times \{b'\})=\{(b,b')\}$,
and some tuple $(a,a')\in R^{\aest}$. In this case we add $(w^i_{j+1},w^i_j)$ to
$R^{\pest}$ and set $h(v^i_j)=a'$ and $h(v^i_{j+1})=a$. The cases $(c)$ and $(d)$
are dealt with in a similar fashion.

By construction, $h$ defines a homomorphism from $\pest$ to $\aest$. We 
shall now prove that $\pest$ is not homomorphic to $\best$. Towards a
contradiction let us assume that there exists such a homomorphism $f$ and
let $b_i$ be the image of $w$ according to $f$. Let us consider the
nodes in $\pest$, $w=w^i_1,w^i_2,\dots,w^i_{l_i}$ associated to the 
path $v^i_1,v^i_2,\dots,v^i_{l_1}$. The first arc $(v^i_1,v^i_2)$ of the path
has been added due to one of the conditions $(a-d)$. If it was added due
to (a) then there exists some predicate $R$
in $\vocab$, some $b,b'\in B$ such that $R^{\best}\cap(\{b\}\times B)=\{(b,b')\}$.
Furthermore $(w^i_1,w^i_2)$ belongs to $R^{\pest}$. Consequently, if $f$
is a homomorphism it must map $w^i_2$ to $b'$. We can do a similar reasoning
for the other conditions $(b-c)$ and for the remaining nodes of the path until
we reach the last element of the path $v^i_{l_i}$. Here we should consider two
cases. If the arc $(v^i_{l_i-1},v^i_{l_i})$ has been added due to conditions
$(a)$ or $(b)$, then can conclude that $w^i_{l_i}$ is $w$ and $f(w)=b_j$ for some $j\neq i$,
obtaining a contradiction.
Otherwise, If the arc $(v^i_{l_i-1},v^i_{l_i})$ has been added due to conditions
$(c)$ or $(d)$, then we can conclude that $f$ is not a homomorphism, getting again a contradiction.

It only remains to show that $\pest$ has pathwidth at most $(2,3)$. The key observation here is
that each one of paths $v^i_1,\dots v^i_{l_i}$ gives rise to a collection of nodes
$w^i_1,\dots,w^i_{l_i}$ of $P$ such that the restriction ${\pest}_{|\{w^i_1,\dots,w^i_{l_i}\}}$
of $\pest$ to $\{w^i_1,\dots,w^i_{l_i}\}$ has the following path-decomposition: 
$\{w^i_1,w^i_2\},\dots,\{w^i_{l_i-1},w^i_{l_i}\}$. Consequently, since $w$ is the only 
common node to each such restriction we can easily conclude that ${\pest}$ has the
path-decomposition: $$\{w,w^1_1,w^1_2\},\{w,w^1_2,w^1_3\},\dots,\{w,w^1_{l_1-1},w^1_{l_1}\},
\{w,w^2_1,w^2_2\},\dots,\{w,w^m_{l_m-1},w^m_{l_m}\}$$ 
\endproof

Using a very similar line of reasoning is it possible to push this results a bit further. In particular
it is possible to show~\cite{Dalmau:2002} that every structure $\best$ that contains only relations invariant under
an operation in the pseudovariety generated by all dual discriminator operations has also
bounded path duality. A complete presentation of this result would require a lengthy introduction
of the algebraic approach to CSP. Since this would lead us out of the scope of this paper
we omit it and instead we refer to~\cite{Dalmau:2002}.

Another related family of constraint satisfaction problems solvable in NL is the class of
all $\csp(\best)$ where $\best$ is a poset with constants invariant under a near-unanimity 
operation~\cite{Krokhin/Larose:stacs03}. Again for space limitations we shall not present this result.
In~\cite{Krokhin/Larose:stacs03} it is shown that
for every such poset $\best$, $\neg\csp(\best)$ is definable in pos FO+TC, that is, the fragment
of FO+TC in which negation and universal quantification is not allowed. It is known (folklore)
that pos FO+TC and linear Datalog have the same expressive power. Consequently, we can
infer that $\best$ has bounded path duality.

\subsection{Implicative Hitting-Set Bounded}

The class of Implicative Hitting-Set Bounded was introduced in~\cite{Creignou/Khanna/Sudan:2001}. Let $k$
be any integer greater than $1$. 
A Boolean relation $R$ is in $k$-IHS-B$+$ if it can be expressed as a CNF where
each clause is of the form $\neg v$, $\neg v\vee w$ or $w_1\vee\cdots\vee w_k$ (here
we do not require that all $w_i$'s are different). Similarly, $R$ is
in $k$-IHS-B$-$ if it can be expressed as a CNF where
each clause is of the form $w$, $\neg v\vee w$ or $\neg v_1\vee\cdots\vee \neg v_k$. A relational
structure $\best$ is implicative Hitting-Set Bounded there exists some $k\geq 2$ such that
all its relations are in $k$-IHS-B$+$ or $k$-IHS-B$-$. It is well known~\cite{Creignou/Khanna/Sudan:2001}
that for every Implicative Hitting-Set Bounded structure $\best$, $\csp(\best)$ is solvable
in NL. We shall prove that it has bounded path width duality.

\begin{lemma}
Let $\best$ be a relational structure containing only relations in $k$-IHS-B$+$ for some $k\geq 2$. Then $\best$
has $(k,k-1+\arity(\best))$-path duality. Similarly, if $\best$ contains only relations in $k$-IHB-B- then $\best$ 
has $(k,k-1+\arity(\best))$-path duality.
\end{lemma}
\proof
We shall consider only the case in which all relations of $\best$ are in $k$-IHS-B$+$. The case $k$-IHS-B$-$
is completely symmetric. We need to introduce a bit of notation. Recall that given a relation
$T$ and a subset $I$ of $\{1,\dots,\arity(T)\}$, $T_{|I}$ denotes the restriction of $T$ to $I$.
We generalize slightly this definition by allowing sequences of integers instead of merely sets. More formally, let $i_1,\dots,i_k$ be 
(not necessarily different) integers in $\{1,\dots,\arity(T)\}$. By $T_{|i_1,\dots,i_k}$ we
denote the $k$-ary relation 
$$\{(a_{i_1},\dots,a_{i_k}) : (a_1,\dots,a_{\arity(T)})\in T\}$$ 

Let us start by introducing a property of unsatisfiable formulas which
will be of most help in providing intuition on the proof. 

\begin{lemma}~\cite{Creignou/Khanna/Sudan:2001}
\label{le:cnf}
Let $\Phi$ be a CNF 
formula containing only clauses of the form $\neg v$, $\neg v\vee w$ or $w_1\vee\cdots\vee w_k$.
If $\Phi$ is unsatisfiable then there exists a clause of the form $w_1\vee\cdots\vee w_k$ such that
for each $i\in\{1,\dots,k\}$ there exists a sequence of variables $w_i=v^i_1,\dots,v^i_{l_i}$
such that $\neg{v^i_{l_i}}$, an $(\neg v^i_j\vee v^i_{j+1})$ for all $j\in\{1,\dots,l_i-1\}$ are  in $\Phi$ 
\end{lemma}

In order to prove the theorem we need a generalization of this result. Let $\aest$ be 
an instance of $\csp(\best)$ and let $h$ be a partial mapping from the universe $A$ 
to $B$. In what follows we shall denote a partial mapping as $(a_1\rightarrow b_1,\dots,a_m\rightarrow b_m)$
where $\{a_1,\dots,a_m\}$ is the domain of the partial mapping and for each $1\leq i\leq m$, $a_i$ is map to $b_i$.
We say that a partial mapping is {\em forbidden} if there exists a predicate $R$ in the vocabulary $\vocab$ of
$\best$ and some integers $i_1,\dots,i_m$ such that $(a_1,\dots,a_m)\in R^{\aest}_{|i_1,\dots,i_m}$ and
$(b_1,\dots,b_m)\not\in R^{\best}_{|i_1,\dots,i_m}$.

\begin{lemma}
\label{le:a}
Let $\aest$ be any unsatisfiable instance of $\csp(\best)$ where $\best$ contains only 
relations in $k$-IHS-B$+$. Then there exists some elements $a^i_j$, $1\leq i\leq k$, $1\leq j\leq l_i$
of $A$ such that $(a^1_1\rightarrow 0,\dots,a^k_1\rightarrow 0)$, $(a^i_{l_i}\rightarrow 1)$, $1\leq i\leq k$,
and $(a^i_j\rightarrow 1,a^i_{j+1}\rightarrow 0)$, $1\leq i\leq k$, $1\leq j\leq l_i-1$ are forbidden partial
mappings
\end{lemma}

\proof
The proof of this lemma is rather straightforward. We only present here an sketch
of the proof. In a first step, taking into account
that every relation in $\best$ is in $k$-IHS-B$+$, it is possible to construct from
$\aest$ and $\best$, an unsatisfiable CNF formula in which the variables are elements of $A$
and every clause is of the form $\neg v$, $\neg v\vee w$, or $w_1\vee\cdots\vee w_k$. Observe
that we can associate to each clause $l_1\vee\cdots\vee l_r$ in $\Phi$ a forbidden partial mapping 
$(v_1\rightarrow b_1,\dots,v_r\rightarrow b_r)$ where $v_i$ is the variable underlying literal $l_i$
and $b_i$ is $0$ if $l_i$ is positive and $1$ otherwise. Finally the result follows by applying 
Lemma~\ref{le:cnf} to formula $\Phi$. 
\endproof

We are now in a position to construct a $\vocab$-structure $\pest$ of $(k,k-1+\arity(\best))$
that is homomorphic to $\aest$ but not homomorphic to $\best$. The universe
$P$ of $\pest$ contains for each $1\leq i\leq k$ and for each element $a^i_j$, $1\leq j\leq l_i$
in the sequence guaranteed by the previous lemma, an element $c^i_j$. If the same element
appears more than once then we make different copies. The universe of $P$ contains also more elements
that will be added as needed.

We now construct the relations of $\pest$.  At the same time we shall
define a homomorphism $h$ from $\pest$ to $\aest$. The intuition of the construction is very
similar to the proof of Lemma~\ref{le:implicational2}.

Let $(a^1_1\rightarrow 0,\dots,a^k_1\rightarrow 0)$ be the first of the forbidden partial mappings guaranteed
to exist by Lemma~\ref{le:a}. Consequently there exists some predicate symbol $R$ in $\vocab$,
some tuple $(v_1,\dots,v_{\arity(R)})\in R^{\aest}$ and 
some $j_1,\dots,j_k\in\{1,\dots,\arity(R)\}$ such that $v_{j_1}=a^1_1,\dots,v_{j_k}=a^k_1$.
Then we include in $R^{\pest}$ a tuple $(w_1,\dots,w_{\arity(R)})$ containing $c^i_1$ in
its $j_i$th position, $i\in\{1,\dots,w_k\}$ and new variables not occurring in $P$ elsewhere.
We set
$h(w_l)$ to be $v_l$. We associate to the forbidden partial mapping $(a^1_1\rightarrow 0,\dots,a^k_1\rightarrow 0)$
a set $S_{(a^1_1\rightarrow 0,\dots,a^k_1\rightarrow 0)}$ that contains all variables $\{w_1,\dots,w_\arity(R)\}$
(we shall use this set later to define the path decomposition of $\pest$).
We proceed in a similar fashion for each forbidden partial mapping guaranteed by Lemma~\ref{le:a}.

By construction $h$ defines a homomorphism from $\pest$ to $\aest$. Furthermore $\pest$ has
has pathwidth at most $(k,k-1+\arity(\best))$ as it is certified by the following path-decomposition:

$$\begin{array}{l}
S_{(a^1_1\rightarrow 0,\dots,a^k_1\rightarrow 0)},S_{(a^1_1\rightarrow 1,a^1_2\rightarrow 0)}\cup\{c^2_1,\dots,c^k_1\},
\dots,S_{(a^1_{l_1-1}\rightarrow 1,a^1_{l_1}\rightarrow 0)}\cup\{c^2_1,\dots,c^k_1\}, \\
S^{(a^1_{l_i}\rightarrow 1)}\cup\{c^2_1,\dots,c^k_1\},S^{(a^2_1\rightarrow 1,a^2_2\rightarrow 0)}\cup\{c^3_1,\dots,c^k_1\},
\dots,S_{(a^k_{l_k-1}\rightarrow 1,a^k_{l_k}\rightarrow 0)},S_{(a^k_{l_k}\rightarrow 1)}
\end{array}$$

Finally, we shall now see that $\pest$ is not homomorphic
to $\best$. Towards a contradiction, let $f$ be such homomorphism. By the construction of $\pest$ we can conclude
that $(c^1_1\rightarrow 0,\dots,c^k_1\rightarrow 0)$ is a forbidden mapping of $\pest$ and $\best$. Consequently,
for some $i$, $f(c^i_1)=1$. Consider now the sequence $c^i_1,c^i_2,\dots,c^i_{l_i}$. By the construction of
$\pest$ for each $j\in\{1,\dots,l_i-1\}$, $(c^i_j\rightarrow 1,c^i_{j+1}\rightarrow 0)$ is a forbidden mapping. Consequently
we can infer by induction that for every $j\in\{1,\dots,l_i\}$, 
$f(c^i_j)=1$. We get then a contradiction with the fact that $(c^i_{l_i}\rightarrow 1)$ is a forbidden
mapping.

\endproof

\subsection{New problems in NL}

In this section we shall use the notion of bounded path as a tool to identify some other
constraint satisfaction problems in NL. We have mentioned already in Section~\ref{sec:implicational}
that every structure invariant under an operation in the pseudovariety generated by
all dual discriminator operations has bounded path duality. 
More examples of constraint satisfaction problems $\csp(\best)$, such that $\best$ has bounded path duality can be found in the  literature about ${\bf H}$-coloring which can be reformulated as the subcase of the general constraint satisfaction problem 
$\csp({\bf H})$ when ${\bf H}$ is a (di)graph. As mentioned in the introduction, the notion of bounded tree duality which is 
intimately related to the notion of path duality (and in fact, inspired it) has been deeply investigated in the field
of ${\bf H}$-coloring. Let us recall that a (di)graph ${\bf H}$ has has bounded tree duality if $\csp(\best)$ has an obstruction
set containing only graphs of treewidth at most $k$ for some fixed $k$. Most of the tractable cases of the ${\bf H}$-coloring
posses bounded tree duality. Indeed, a closer inspection of those results shows that most times, the obstruction set corresponding to
a given (di)graph ${\bf H}$ contains only structures with bounded pathwidth. Consequently, as a direct consequence
of Proposition~\ref{pro:NL}, we can lower the complexity of this problems from P to NL with virtually no effort.

Let us revisit now some of these examples. For undirected graphs, the only case solvable in polynomial time 
corresponds to bipartite graphs~\cite{Hell/Nesetril:1990}. It is fairly easy to see that the set of all odd cycles
is an obstruction set of $\csp({\bf H})$ for any bipartite graph ${\bf H}$. As cycles have pathwidth at most $(2,3)$ we
can conclude that every bipartite graph ${\bf H}$ has $(2,3)$-path duality. We have thus strengthened the dichotomy
result of~\cite{Hell/Nesetril:1990}: for every undirected graph ${\bf H}$, $\csp({\bf H})$ is in NL or NP-complete.

For directed graphs, it is a direct consequence of the results in~\cite{Hell/Zhu:1994,Hell/Zhu:1995} 
(see also~\cite{Hell/Nesetril/Zhu:1996}) that
every oriented path, directed cycle, and even more generally, every unbalanced oriented cycle has bounded path duality. 

Let us present these results.
An oriented path is a digraph ${\bf H}$ with nodes $p_0,\dots,p_n$ such that for 
each $i\in\{0,\dots,n-1\}$ either $(p_i,p_{i+1})$ or $(p_{i+1},p_i)$ is an edge of ${\bf H}$
and it does not contain any other edge. It was
shown in~\cite{Hell/Zhu:1994} that for every oriented path ${\bf H}$, $\csp({\bf H})$ has an obstruction
set containing only oriented paths. It is easy to observe that the Gaifman graph of an oriented path is 
a path and hence has pathwidth at most $(1,2)$. Consequently, oriented paths have $(1,2)$-path duality.

An oriented cycle is a digraph $H$ with nodes $p_0,\dots,p_n$ such that
for each $i\in\{0,\dots,n\}$, either $(p_i,p_{i+1})$ or $(p_{i+1},p_i)$ is an edge of ${\bf H}$ 
(the sum is modulo $n+1$). If additionally all edges are in the same direction then ${\bf H}$ is
called a directed path. More generally, if the number of forward edges is different than the number
of backward edges then ${\bf H}$ is called unbalanced. 

In~\cite{Hell/Zhu:1995}, it is shown that
for each unbalanced oriented cycle ${\bf H}$, $\csp({\bf H})$ has an obstruction set containing only oriented
cycles. As oriented cycles have pathwidth at most $(2,3)$ we can conclude that ${\bf H}$ has $(2,3)$-path duality.

We conclude this section by examining the problem of deciding whether a given structure ${\best}$ has bounded path duality.
In general, this problem is not known to be decidable.
However, in a particular case it is possible to show decidability, by mimicking 
some results originally proven in~\cite{Feder/Vardi:1998}. 

\begin{theorem}
\label{the:1k}
For every $1\leq k$, the problem of deciding whether a $\vocab$-structure $\best$ has $(1,k)$-path duality is decidable.
\end{theorem}
\proof
By Theorem~\ref{the:expressivitygame}, there exists a sentence $\varphi$ in $1$-adic (also called monadic) monotone SNP such that
for every $\vocab$-structure $\aest$, $\aest\models\varphi$ iff the Duplicator has a winning strategy for
the $(1,k)$-PR game on $\aest$ and $\best$. It is widely known (see~\cite{Feder/Vardi:1998}) that there exists a 
sentence $\psi$ in monadic monotone SNP such that for every $\vocab$-structure $\aest$, $\aest\models\psi$ iff
$\aest\in\csp(\best)$. Thus deciding $(1,k)$-path duality is equivalent to decide whether $(\varphi\implies\psi)$
is a tautology for finite structures. The latter is decidable since so is containment for monadic monotone SNP~\cite{Feder/Vardi:1998}.
\endproof

\section{Open Problems}

It is unknown whether bounded path duality captures the class of CSPs solvable in NL, i.e, whether there exists
a structure that does not have bounded path duality such that $\csp(\best)$ is solvable in NL.
In this direction, it is not difficult to show, using some results in~\cite{Gradel:1992},
that for every structure $\best$, if $\csp(\best)$ is in NL, then it can be expressed in restricted Krom monadic SNP with 
equality, $s(x,y)$ (true when $y$ is the immediate successor of $x$ in some total ordering),
$0$ and {\em max} (first and last elements of the total ordering respectively). Here we understand
that the homomorphism condition only applies to non-built-in predicates. 

Another interesting problem is to extend the class of structures $\best$ known to have bounded path duality
via closure conditions. With that respect, it would be interesting to decide whether majority operations,
or even more generally, near-unanimity operations~\cite{Szendrei:1987} (see also~\cite{Jeavons/Cohen/Cooper:1998}) 
guarantee bounded path duality. This result is true, in particular, for structures with domain of size at most $2$.

\section{Acknowledgments}

I am grateful to Phokion Kolaitis for stimulating discussions and for sharing with me his expertise on
logic and games.

\bibliographystyle{plain}
\bibliography{../../../common/bibliography}

\end{document}